\documentclass[12pt]{iopart}
\usepackage{iopams}
\usepackage{epsfig}

\def\ket#1{\mathinner{|{#1}\rangle}}
\def\braket#1{\mathinner{\langle{#1}\rangle}}

{\catcode`\|=\active
  \gdef\Braket#1{\left<\mathcode`\|"8000\let|\bravert {#1}\right>}}
\def\bravert{\egroup\,\vrule\,\bgroup}

\newcommand{\sch}{Schr\"{o}dinger }
\newcommand{\za} {Zeldovich approximation }

\newcommand{\wm} {wave-mechanical }
\newcommand{\fp} {free-particle }
\newcommand{\ada}{adhesion approximation }
\newcommand{\nb} {$N$-body }
\newcommand{\n}{\noindent}
\newcommand{\sps}{Schr\"{o}dinger-Poisson }

\begin{document}

\title{Gravitational instability via the Schr\"{o}dinger equation}
\author{C J Short and P Coles}
\address{Cripps Centre for Astronomy and Particle Theory, School of Physics and Astronomy, University of Nottingham, University Park, Nottingham, UK, NG7 2RD}
\eads{\mailto{ppxcjs@nottingham.ac.uk}, \mailto{peter.coles@nottingham.ac.uk}}

\begin{abstract}
We explore a novel approach to the study of large-scale structure
formation in which self-gravitating cold dark matter (CDM) is represented by
a complex scalar field whose dynamics are governed by coupled \sch and Poisson
equations. We show that, in the quasi-linear regime, the \sch equation
can be reduced to the free-particle \sch equation. We advocate using the \fp
\sch equation as the basis of a new approximation method - the {\it \fp
approximation} - that is similar in spirit to the successful adhesion model. In
this paper we test the \fp approximation by appealing to a planar
collapse scenario and find that our results are in excellent agreement with
those of the Zeldovich approximation, provided care is taken when choosing a
value for the effective Planck constant in the theory. We also discuss how
extensions of the \fp approximation are likely to require the inclusion of a
time-dependent potential in the \sch equation. Since the \sch equation with a
time-dependent potential is typically impossible to solve exactly, we
investigate whether standard quantum-mechanical approximation techniques can
be used, in a cosmological setting, to obtain useful solutions of the \sch
equation. In this paper we focus on one particular approximation method:
time-dependent perturbation theory (TDPT). We elucidate the
properties of perturbative solutions of the \sch equation by considering a
simple example: the gravitational evolution of a plane-symmetric density
fluctuation. We use TDPT to calculate an approximate solution of the relevant
\sch equation and show that this perturbative solution can be used to
successfully follow gravitational collapse beyond the linear regime, but there
are several pitfalls to be avoided.
\end{abstract}

\submitto{\it{J. Cosmol. Astropart. Phys.}}

\section{Introduction}

The local universe displays a rich hierarchical pattern of galaxy
clustering that encompasses a vast range of length scales,
culminating in rich clusters, super-clusters and filaments. However, the early
universe was almost homogeneous with only slight temperature fluctuations seen
in the cosmic microwave background radiation. Models of structure formation
link these observations through the effect of gravity, relying on the fact that
small initially over-dense regions accrete additional matter as the universe
expands (a mechanism known as gravitational instability). The growth of
density perturbations via gravitational instability is well understood in the
linear regime, but the non-linear regime is much more complicated and
generally not amenable to analytic solution. Numerical \nb
simulations have led the way towards an understanding of strongly developed
clustering. Although such calculations have been priceless in establishing
quantitative predictions of the large-scale structure expected to arise in a
particular cosmology, it remains important to develop as full an
analytical understanding as possible. After all, simulating a thing
is not quite equivalent to understanding it.

Analytical methods for studying the evolution of cosmological density
perturbations fall into two broad classes (for a review, see \cite{sahni} and references therein). First there are techniques based on applying
perturbation theory (PT) to a hydrodynamical description of self-gravitating
cold dark matter (CDM). These Eulerian approaches range from simple first-order
(linear) PT (e.g. \cite{peeb}) through to higher-order approaches of vastly
increased complexity (e.g. \cite{cat1, bern}). First-order Eulerian PT (the
so-called {\it linearized fluid approach}) has been the
mainstay of structure formation theory for many decades, by virtue of its
simplicity and its robustness on large scales where density fluctuations are very much smaller than the mean density. When extrapolated to
smaller scales, it provides a useful indicative measure of clustering
strength, but can lead to absurdities (such as a negative matter density)
if taken too far. Alternatively, there are Lagrangian approaches in which the
trajectories of individual CDM particles are perturbed, rather than
macroscopic fluid quantities (e.g. \cite{buc1, buc2, bouch, cat2}). Even
first-order Lagrangian PT, the celebrated {\it Zeldovich approximation}
\cite{zel}, is capable of following the gravitational collapse of density
perturbations into the quasi-linear regime, beyond the breakdown of linear Eulerian PT. Comparisons with full \nb
calculations have shown that the \za can successfully describe the
quantitative morphology of the clustering pattern (e.g. \cite{coles2}), as
long as particle trajectories do not intersect. When particle trajectories
cross (a phenomenon known as {\it shell-crossing}) the density field develops
a formal singularity (known as a {\it caustic}). Since the \za is
purely kinematical, particles simply pass through the caustic and continue
along their original trajectories. This leads to large-scale structure being
rapidly `washed out' in a manner not observed in \nb simulations. A simple
extension of the \za that overcomes this problem is the {\it adhesion
  approximation} \cite{gurb}. In the adhesion approximation, particles move
according to the \za until trajectories intersect, but are assumed to `stick'
to each other when shell-crossing occurs. This sticking is achieved by
including an artificial viscosity term in the equations of motion to mimic
the action of the strong gravitational forces acting in the vicinity of a
caustic. The adhesion model has proved very successful in explaining the
pattern of large-scale structure observed in the universe; comparisons with \nb
simulations have shown impressive agreement far into the non-linear regime
(e.g. \cite{kof1, wein, nuss, kof2, mell}). 

In this paper we explore a radically different approach to the study of
large-scale structure formation based on a \wm description of
self-gravitating CDM. Following a suggestion by Widrow and Kaiser
\cite{widrow}, we transform the usual hydrodynamical equations of motion into
a non-linear \sch equation coupled to a Poisson equation describing Newtonian
gravity. The non-linear \sch equation is similar to the familiar
linear \sch of quantum mechanics, except for the presence of an extra term
known as the {\it quantum pressure}. The reasoning behind this transformation is not
new - in essence it dates back to Madelung \cite{madel} - but it is quite neglected in cosmology. Coles \cite{coles} showed that the \wm approach provides a
natural explanation of why the distribution function of density fluctuations,
evolved from Gaussian initial conditions, should be close to the log-normal
form that is observed. By studying the evolution of a one-dimensional
sinusoidal density perturbation, Coles and Spencer \cite{coles3} demonstrated
that the \wm approach fares well in comparison with the Zeldovich
approximation up until particle trajectories cross and that, when
shell-crossing occurs, the \wm density field remains non-singular. Their work
relied exclusively on numerical solutions of the \sch equation; our
objective here is to investigate how the \sps system can be solved by using
approximation techniques instead. We show that, in an expanding
universe, the \sch equation can be reduced to the exactly solvable
`free-particle' \sch equation by utilizing results from first-order Eulerian
and Lagrangian PT. We advocate using the \fp \sch equation as the basis of a
new approximation method - the {\it \fp approximation} - that promises to be
capable of evolving cosmological density perturbations into the quasi-linear
regime. As we shall see, the \fp approximation can essentially be thought of
as an alternative to the adhesion model in which the viscosity term is replaced by the quantum pressure term. The performance of
the new \fp approximation is assessed, relative to the Zeldovich
approximation, by considering a simple example of gravitational
collapse. Particular attention is paid to elucidating the effect of the
quantum pressure term since this has not been investigated previously. We also
discuss how extensions of the \fp approximation based on higher-order PT are
likely to lead to a \sch equation with a time-dependent potential that cannot
be solved exactly. Consequently, we feel it is important to ascertain whether
standard quantum-mechanical approximation techniques can be used, in a
cosmological context, to obtain solutions of the \sch equation with a
time-dependent potential. In this work we concentrate on one particular
approximation method: time-dependent perturbation theory (TDPT). Our goal is
to investigate the properties of perturbative solutions of the \sch equation
in a cosmological setting; we perform our investigation by appealing to a
simple gravitational collapse scenario. Although we restrict our attention to
idealized examples in this work, the results we obtain will remain pertinent
when the \wm method is applied to more general structure formation problems in
the future.

The outline of this paper is as follows: First, in order to keep the
paper as self-contained as possible, we briefly outline the basic theory of
cosmological structure formation including the standard fluid approach, the
Zeldovich approximation, the adhesion model, the \wm approach in general and
the \fp approximation in particular. Next, in section \ref{egcoll}, we
describe how the gravitational collapse of a plane-symmetric sinusoidal
density perturbation can be modelled by using (i) the \fp approximation and
(ii) a perturbative solution of the \sch equation with a time-dependent potential. We present our results in section \ref{results} and conclude in section \ref{conc}.

\section{Cosmological structure formation}

Observations suggest that there is approximately five times as much
non-baryonic CDM in the universe as there is ordinary (luminous)
matter. Consequently, large-scale structure formation is commonly studied by
considering the gravitational amplification of fluctuations in the CDM
distribution only. In standard treatments, collisionless CDM is assumed to be
an ideal fluid with zero pressure (so-called {\it dust}) and the evolution
equations for perturbations in the CDM fluid are obtained by
linearizing the Einstein field equations about an expanding homogeneous and
isotropic Friedmann-Robertson-Walker (FRW) cosmology (e.g. \cite{bard, kodama,
  mukh}). However, in this paper we will
assume that the length scale of the perturbations is much smaller than the
Hubble radius so that a Newtonian treatment is adequate (e.g. \cite{peeb}). 

\subsection{The fluid approach}

In an expanding universe, the Newtonian dynamical equations governing the
evolution of fluctuations in a fluid of collisionless self-gravitating CDM can be written in the form:

\begin{eqnarray}
\frac{\partial\mathbf{v}}{\partial t}+H\mathbf{v}+\frac{1}{a}(\mathbf{v}\cdot\nabla_{\mathbf{x}})\mathbf{v}+\frac{1}{a}\nabla_{\mathbf{x}}\Phi=0,\label{eul}\\
\frac{\partial\delta}{\partial t}+\frac{1}{a}\nabla_{\mathbf{x}}\cdot[(1+\delta)\mathbf{v}] = 0,\label{cty}\\
\nabla_{\mathbf{x}}^2 \Phi-4\pi Ga^2\rho_{\rm b,c}\delta=0,\label{poiss}
\end{eqnarray}

\n where $t$ is cosmological proper time and $\mathbf{x}=\mathbf{x}(t)$ are
comoving coordinates, related to physical coordinates $\mathbf{r}$ via
$\mathbf{r}=a\mathbf{x}$. Here the scale factor $a=a(t)$ has been normalized
so that its value at the present epoch $t_0$ is $a_0=a(t_0)=1$. The Hubble
parameter $H=H(t)$ is defined by $H\equiv\dot{a}/a$ where a dot denotes a
derivative with respect to $t$. The peculiar velocity field
$\mathbf{v}=\mathbf{v}(\mathbf{x},t)$ is given by
$\mathbf{v}=a\dot{\mathbf{x}}$ and the potential $\Phi=\Phi(\mathbf{x},t)$ is the peculiar Newtonian gravitational potential. The density contrast
$\delta=\delta(\mathbf{x},t)$ is $\delta=\rho/\rho_{{\rm b,c}}-1$ where
$\rho=\rho(\mathbf{x},t)$ is the CDM density field and $\rho_{{\rm
    b,c}}=\rho_{{\rm b,c}}(t)$ is the CDM density in the homogeneous FRW background.

It is straightforward to show (e.g. \cite{peeb}) from the fluid
equations (\ref{eul}), (\ref{cty}) and (\ref{poiss}) that, to first-order in Eulerian PT, density perturbations in the CDM fluid grow according to
$\delta=D\delta_{\rmi}$ where $\delta_\rmi$ is the density contrast at some
initial time $t_\rmi$ and the linear growth factor $D=D(t)$ is the growing mode solution of

\begin{equation}
\label{lg}
\ddot{D}+2H\dot{D}-4\pi G\rho_{\rm b,c}D=0,
\end{equation}

\n normalized so that $D_\rmi=D(t_\rmi)=1$. The linear growth factor can then be used to write the fluid equations (\ref{eul}), (\ref{cty}) and (\ref{poiss}) in the alternative form:

\begin{eqnarray}
\frac{\partial\mathbf{u}}{\partial
  D}+(\mathbf{u}\cdot\nabla_{\mathbf{x}})\mathbf{u}+\frac{3\Omega_{\rm
  c}}{2f^2 D}\mathbf{u}+\nabla_{\mathbf{x}}\Theta=0,\label{Deul}\\
\frac{\partial\delta}{\partial D}+\nabla_{\mathbf{x}}\cdot[(1+\delta)\mathbf{u}] = 0,\label{Dcty}\\
\nabla_{\mathbf{x}}^2 \Theta - \frac{3\Omega_{\rm c}}{2f^2 D^2}\delta=0,\label{Dpoiss}
\end{eqnarray}

\n where the function $f=f(D)$ is given by $f\equiv\dot{D}/HD$ and
$\Omega_{\rm c}=\Omega_{\rm c}(D)$ is the familiar CDM density parameter:
$\Omega_{\rm c}=8\pi G\rho_{\rm b,c}/3H^2$. The gravitational potential
$\Theta=\Theta(\mathbf{x},D)$ is related to the peculiar gravitational
potential $\Phi$ via $\Theta=\Phi/a^2\dot{D}^2$. Similarly, the comoving
velocity field $\mathbf{u}=\mathbf{u}(\mathbf{x},D)$, defined by
$\mathbf{u}\equiv\rmd\mathbf{x}/\rmd D$, is obtained from the peculiar velocity
$\mathbf{v}$ by a simple rescaling: $\mathbf{u}=\mathbf{v}/a\dot{D}$. It is
well known that linear Eulerian PT implies an irrotational velocity field: $\mathbf{u}=-\nabla_{\mathbf{x}}\phi$, where the velocity potential $\phi=\phi(\mathbf{x},D)$ is related to the gravitational potential $\Theta$ via

\begin{equation}
\label{phirel}
\phi=\frac{2f^2 D}{3\Omega_{\rm c}}\Theta.
\end{equation}

\n The linearized fluid equations only provide an accurate description of
gravitational instability in the linear regime $\delta\ll 1$. A considerably
more powerful method than the linearized fluid approach is the \za \cite{zel}
which is capable of following the evolution of cosmological density perturbations into the quasi-linear regime $\delta\sim 1$.

\subsection{The \za}

The Zeldovich approximation is a Lagrangian approach (formally first-order
Lagrangian PT) in which individual particle trajectories are considered: the
comoving Eulerian coordinate $\mathbf{x}=\mathbf{x}(\mathbf{q},D)$ of a
particle initially located at a comoving Lagrangian coordinate $\mathbf{q}$ is given by

\begin{equation}
\label{za}
\mathbf{x}=\mathbf{q}+(D-1)\mathbf{s},
\end{equation}

\n where $\mathbf{s}=\mathbf{s}(\mathbf{q})$ is a time-independent vector
field. Requiring that mass is conserved immediately leads to the following
expression for the CDM density field in the Zeldovich approximation:

\begin{equation}
\label{delc}
\delta=\frac{(1+\delta_\rmi)}{\mathcal{J}}-1,
\end{equation}

\n where $\delta_\rmi=\delta_\rmi(\mathbf{q})$ is the initial density
perturbation and $\mathcal{J}=\mathcal{J}(\mathbf{q},D)$ is the determinant of
the Jacobian of the mapping (\ref{za}) between Lagrangian and Eulerian
coordinates. The density field (\ref{delc}) is known as the {\it continuity} density. 

The \za can also be formulated in Eulerian space by noting that
(\ref{za}) implies $\rmd\mathbf{u}/\rmd D=0$ along a particle
trajectory, provided particle trajectories do not intersect. Using the definition of the convective derivative $\rmd/\rmd
D=\partial/\partial D+\mathbf{u}\cdot\nabla_{\mathbf{x}}$, it follows that the \za corresponds to

\begin{equation}
\label{eza}
\frac{\partial\mathbf{u}}{\partial D}+(\mathbf{u}\cdot\nabla_{\mathbf{x}})\mathbf{u}=0,
\end{equation}

\n in Eulerian space. Comparing (\ref{eza}) with the Euler equation
(\ref{Deul}) it is apparent that the \za also guarantees an irrotational
velocity field $\mathbf{u}=-\nabla_{\mathbf{x}}\phi$. We can then integrate
(\ref{eza}) to obtain the so-called {\it Zeldovich-Bernoulli} \cite{nuss2} equation:

\begin{equation}
\label{zb}
\frac{\partial\phi}{\partial D}-\frac{1}{2}\left|\nabla_{\mathbf{x}}\phi\right|^2=0,
\end{equation}

\n which possesses an analytical solution $\phi=\phi(\mathbf{x},D)$ of the form

\begin{equation}
\label{zbsol}
\phi=\phi_\rmi-\frac{1}{2}(D-1)\left|\mathbf{s}\right|^2,
\end{equation}

\n where $\phi_\rmi=\phi_\rmi(\mathbf{q})$ is the initial velocity
potential and $\mathbf{x}$ and $\mathbf{q}$ are related by the mapping
(\ref{za}). As in the linearized fluid approach, the velocity potential $\phi$
is related to the gravitational potential $\Theta$ via (\ref{phirel}) and thus the Poisson equation (\ref{Dpoiss}) implies 

\begin{equation}
\label{deld}
\delta=D\nabla_{\mathbf{x}}^2\phi,
\end{equation}

\n which is known as the {\it dynamical} density field. In general, the
continuity and dynamical density fields are not equivalent. However, in one
dimension, the density fields (\ref{delc}) and (\ref{deld}) coincide and the
\za is an exact solution of the fluid equations as long as there is no
shell-crossing (for a discussion, see \cite{sahni} and references therein).

The \za breaks down when particle trajectories cross since the mapping
(\ref{za}) from $\mathbf{q}$ to $\mathbf{x}$ is no longer unique. At a point
where particle trajectories intersect, the velocity field becomes multi-valued,
$\mathcal{J}\equiv 0$ and the density
field develops a formal singularity. The strong gravitational forces acting in
the vicinity of the caustic should cause particles to be pulled towards
it. However, since the \za is kinematical, particles simply pass
straight through the caustic and structure is rapidly `smeared out' in an
unphysical manner. The \ada \cite{gurb} was specifically designed to tackle
this shortcoming of the Zeldovich approximation.

\subsection{The \ada}

In the adhesion approximation, particles follow Zeldovich trajectories until
shell-crossing occurs. However, when particle trajectories cross, the
particles are assumed to `stick' to each other. As a result, the singularities
predicted by the \za are regularized and stable structures are formed, rather
than being `washed out'. Mathematically, the adhesion model is obtained from
the \za by including an artificial viscosity term in (\ref{eza}) so that:

\begin{equation}
\label{burg}
\frac{\partial\mathbf{u}}{\partial D}+(\mathbf{u}\cdot\nabla_{\mathbf{x}})\mathbf{u}-\mu\nabla_{\mathbf{x}}^2\mathbf{u}=0,
\end{equation}

\n which is known as {\it Burgers' equation} \cite{burg}. The constant $\mu>0$
has dimensions of $L^2$ and can be thought of as a viscosity coefficient. A
forcing term of the form $\nabla_{\mathbf{x}}\eta$, where
$\eta=\eta(\mathbf{x},D)$ is a random potential, can be included in Burgers'
equation to extend the adhesion model to the intergalactic medium
(e.g. \cite{matt}). In the cosmologically relevant case of an irrotational velocity field $\mathbf{u}=-\nabla_{\mathbf{x}}\phi$, (\ref{burg}) can be integrated to obtain

\begin{equation}
\label{kpz}
\frac{\partial\phi}{\partial D}-\frac{1}{2}\left|\nabla_{\mathbf{x}}\phi\right|^2-\mu\nabla_{\mathbf{x}}^2\phi=0.
\end{equation}

\n From a practical point of view, an attractive feature of Burgers' equation
(or, equivalently, (\ref{kpz})) is that it possesses an analytic solution. In the special case $\mu\rightarrow 0$ (known as the {\it
  inviscid} limit) a geometrical interpretation of the solution of Burgers'
equation can be used to determine the `skeleton' of the large-scale structure
present at any given time (e.g. \cite{gurb, kof1, kof2}). In this
limit the structures formed in the adhesion model are infinitely thin and
the \ada reduces exactly to the \za outside of mass concentrations. For finite
values of $\nu$, the viscosity term has an effect away from regions where
particle trajectories cross and causes density perturbations to be suppressed
on scales $\lesssim\mu^{1/2}$ (e.g. \cite{wein}). The adhesion and Zeldovich
approximations are then no longer identical outside of collapsing regions
although, depending on the actual value of $\nu$, they become similar at a certain distance.

The success of the adhesion model, relative to \nb calculations, has led to
adhesive gravitational clustering becoming an important concept in the study
of cosmological structure formation. Although the original motivation for the
\ada was purely {\it phenomenological}, it has recently been shown that, under
certain simplifying assumptions, Burgers' equation can be naturally derived
from the coarse-grained equations of motion \cite{buch1, buch2, dom1, dom2,
  buch3}. In these modern interpretations of the adhesion model, the regularizing parameter $\mu$ emerges, not as a (constant) viscosity coefficient, but rather as a (density-dependent) {\it gravitational multi-stream coefficient}, arising from the self-gravitation of a multi-stream system.

\subsection{The wave-mechanical approach}

An alternative approach to the study of large-scale structure formation was
suggested by Widrow and Kaiser \cite{widrow}. They proposed a \wm description
of self-gravitating matter in which CDM is represented by a complex scalar
field $\psi=\psi(\mathbf{x},D)$ whose dynamics are governed by coupled \sch
and Poisson equations (see \cite{widrow2} for a relativistic extension of the
original Newtonian theory). In this section we give a pedagogical derivation
of the \wm formalism in an expanding universe since this has not been
presented in the literature previously.

We begin by assuming an irrotational velocity field
$\mathbf{u}=-\nabla_{\mathbf{x}}\phi$. As discussed previously this is
guaranteed to be the case in the linear and quasi-linear regimes and will
remain so as long as there is no shell-crossing by Kelvin's circulation
theorem. The Euler equation (\ref{Deul}) can then be integrated to obtain the Bernoulli equation

\begin{equation}
\label{bern}
\frac{\partial\phi}{\partial D}-\frac{1}{2}\left|\nabla_{\mathbf{x}}\phi\right|^2-\mathcal{V}=0,
\end{equation}

\n where the effective potential $\mathcal{V}=\mathcal{V}(\mathbf{x},D)$
depends on both the gravitational potential $\Theta$ and velocity potential $\phi$:

\begin{equation}
\label{V}
\mathcal{V}=\Theta-\frac{3\Omega_{\rm c}}{2f^2 D}\phi.
\end{equation}

\n The Bernoulli equation (\ref{bern}) can be combined with the continuity
equation (\ref{Dcty}) by performing a {\it Madelung transformation} \cite{madel}

\begin{equation}
\label{mad}
\psi=(1+\delta)^{1/2}\exp{\left(\frac{-\rmi\phi}{\nu}\right)},
\end{equation}

\n where $\nu$ is a real parameter with
dimensions of $L^2$. The wavefunction $\psi$ provides an elegant description
of both density and velocity fields in a single complex function. Applying
the Madelung transformation leads to the following coupled Schr\"{o}dinger-Poisson system:

\begin{eqnarray}
\rmi\nu\frac{\partial\psi}{\partial D} = \left(-\frac{\nu^2}{2}\nabla_{\mathbf{x}}^{2} + \mathcal{V}+\mathcal{P}\right)\psi,\label{Dtdse}\\
\nabla_{\mathbf{x}}^{2}\left[\mathcal{V}-\frac{3\Omega_{\rm c}}{2f^2
    D}\nu\arg{(\psi)}\right] - \frac{3\Omega_{\rm c}}{2f^2 D^2}\left(|\psi|^{2}-1\right)=0,\label{Dwmpoiss}
\end{eqnarray}

\n where $\mathcal{P}=\mathcal{P}(\mathbf{x},D)$ is the so-called {\it quantum pressure} term, given by

\begin{equation}
\mathcal{P}=\frac{\nu^{2}}{2}\frac{\nabla_{\mathbf{x}}^{2}|\psi|}{|\psi|}.
\end{equation}

\n If we neglect the quantum pressure term for the moment
then we are left with the more familiar linear \sch equation

\begin{equation}
\label{lDtdse}
\rmi\nu\frac{\partial\psi}{\partial D} = \left(-\frac{\nu^2}{2}\nabla_{\mathbf{x}}^{2} + \mathcal{V}\right)\psi.
\end{equation}

\n Inserting the Madelung transformation into (\ref{lDtdse}) yields the usual
continuity equation (\ref{Dcty}) and a modified Bernoulli equation of the form

\begin{equation}
\label{bern2}
\frac{\partial\phi}{\partial D}-\frac{1}{2}\left|\nabla_{\mathbf{x}}\phi\right|^2-\mathcal{V}+\mathcal{P}=0,
\end{equation}

\n so the quantum pressure term appears in the fluid equations. This implies
that we are free to drop the quantum pressure term from the non-linear \sch
equation (\ref{Dtdse}) and include it in the Bernoulli equation instead; this is the approach that we will adopt throughout this paper.

The parameter $\nu$ appears in the \sch equation in a manner analogous to
the Planck constant in real quantum mechanics. However, the system we are
considering here is entirely classical and thus $\nu$ is treated as an
adjustable parameter that controls the quantum pressure term
$\mathcal{P}\propto\nu^2$ appearing in the Bernoulli equation
(\ref{bern2}). The quantum pressure term acts as a regularizing term in the
fluid equations, preventing the generation of multi-stream regions and
singularities in the density field when particle trajectories cross. This was
demonstrated by Coles and Spencer \cite{coles3} who used the \wm approach to follow the gravitational
collapse of a one-dimensional sinusoidal density perturbation through
shell-crossing. The effect of the quantum pressure is thus qualitatively
similar to that of the term $\mu\nabla_{\mathbf{x}}^2\phi$ in the adhesion
model and $\nu$ plays a similar role to the viscosity parameter
$\mu$, in the sense that they are both approximations to a general
gravitational multi-stream coefficient arising in collisionless systems. The
link between the \wm approach and the \ada will be further discussed below. In
the {\it semi-classical} limit $\nu\rightarrow 0$ we expect the effect of the
quantum pressure term to be minimized and thus our \wm representation of
self-gravitating CDM will approach the standard hydrodynamical description.

The main deficiency of the \wm approach described here is that it is based on
the fluid equations and thus implicitly assumes the existence of a single
fluid velocity at each point. Consequently, the Schr\"{o}dinger-Poisson system
is incapable of properly describing the formation of multi-stream
regions. However, this is not an intrinsic problem of the \wm formalism, but
is instead a consequence of the simple Madelung form of the wavefunction we
have used. Widrow and Kaiser \cite{widrow} showed that one can deploy more
sophisticated representations of the wave function, such as the coherent-state
formalism of Husimi \cite{husimi}, that allow for
multi-streaming. The \wm approach can then be adapted to follow the full
Vlasov evolution of the phase-space distribution function beyond the laminar
flow regime \cite{widrow, widrow2, davies}. These extensions are beyond the scope of this paper.

\subsubsection{The \fp approximation}
\label{fpsec}

The full Schr\"{o}dinger-Poisson system cannot be solved analytically and
so we must resort to either numerical or approximation methods; in this work
we are concerned with the latter approach. The Schr\"{o}dinger-Poisson system
can be significantly simplified by using the fact that, to first-order in
Eulerian and Lagrangian PT, the gravitational and velocity potentials are
related via $\phi=2f^2 D\Theta/3\Omega_{\rm c}$. In this case, it is evident
from (\ref{V}) that the effective potential $\mathcal{V}$ is identically zero
and the \sch equation (\ref{lDtdse}) reduces to the exactly solvable \fp \sch equation 

\begin{equation}
\label{fptdse}
\rmi\nu\frac{\partial\psi}{\partial D}=-\frac{\nu^2}{2}\nabla_{\mathbf{x}}^{2}\psi.
\end{equation}

\n Accordingly, the Poisson equation (\ref{Dwmpoiss}) decouples from the \sch
equation and provides the following relationship between the amplitude and
phase of the wavefunction:

\begin{equation}
\label{ampphs}
\nu\nabla_{\mathbf{x}}^{2}\left[\arg{(\psi)}\right] + \frac{1}{D}\left(|\psi|^{2}-1\right)=0.
\end{equation}

\n Inserting the Madelung transformation (\ref{mad}) into (\ref{fptdse}) and (\ref{ampphs}) we obtain 

\begin{equation}
\label{bern3}
\frac{\partial\phi}{\partial D}-\frac{1}{2}\left|\nabla_{\mathbf{x}}\phi\right|^2+\mathcal{P}=0,
\end{equation}

\n along with the continuity equation (\ref{Dcty}) and the familiar relation
(\ref{deld}). The modified Bernoulli equation (\ref{bern3}) looks like the
\ada (\ref{kpz}), except that the term $\mu\nabla_{\mathbf{x}}^2\phi$ has been
replaced by the quantum pressure $\mathcal{P}$. Since the quantum pressure
term cannot be written in a form proportional to $\nabla_{\mathbf{x}}^2\phi$,
we formulate an alternative to the adhesion model based on the \fp
\sch equation (\ref{fptdse}) and the Poisson-like equation (\ref{ampphs}). We
call this new approximation scheme the {\it \fp approximation}. In
\ref{path} we show that, in the semi-classical limit $\nu\rightarrow 0$, the
\fp approximation reduces to the \za prior to shell-crossing. The quantum
pressure term will then only become important in regions where particle
trajectories intersect.

\section{Examples of gravitational collapse using the \sch equation}
\label{egcoll}

\subsection{A simple test of the \fp approximation}
\label{fptst}

The free-particle approximation should be capable of following the evolution of density fluctuations into the quasi-linear regime. In order to verify this hypothesis, we test the \fp approximation by
appealing to a simple example: the gravitational collapse of a plane-symmetric density perturbation. The assumption of plane-symmetry implies that
we are effectively considering a one-dimensional problem. This is advantageous
since, in one dimension, the \za provides an exact solution of the equations
of motion until shell-crossing occurs. The performance of the new \fp
approximation can then be assessed relative to the exact Zeldovich
solution. We will consider the case where the parameter $\nu$ is finite
(rather than $\nu\rightarrow 0$) and investigate the effect of the quantum
pressure term as $\nu$ is varied. This is an important issue to address since
the quantum pressure term is an integral part of the \fp approximation (and,
indeed, the \wm approach in general). For illustrative purposes we assume the plane-symmetric initial density perturbation $\delta_\rmi=\delta_\rmi(x)$ is sinusoidal:

\begin{equation}
\label{idc}
\delta_\rmi=\delta_{\rm a}\cos{\left(\frac{2\pi x}{d}\right)},
\end{equation}

\n where $d$ is the comoving period of the perturbation and $0<|\delta_{\rm
  a}|\leq 1$ to ensure that the initial CDM density field is everywhere non-negative. The corresponding initial velocity potential perturbation is found from (\ref{ampphs}):

\begin{equation}
\label{iphi}
\phi_\rmi = -\left(\frac{d}{2\pi}\right)^2\delta_\rmi.
\end{equation} 

\n The one-dimensional \fp \sch equation

\begin{equation}
\label{1dfptdse}
\rmi\nu\frac{\partial\psi}{\partial D}=-\frac{\nu^2}{2}\frac{\partial^2\psi}{\partial x^2}
\end{equation}

\n must then be solved subject to the initial condition 

\begin{equation}
\label{ipsi}
\psi_\rmi=(1+\delta_\rmi)^{1/2}\exp{\left(\frac{-\rmi\phi_\rmi}{\nu}\right)},
\end{equation}

\n with $\delta_\rmi$ and $\phi_\rmi$ given by (\ref{idc}) and
(\ref{iphi}), respectively. In order to simplify calculations we consider a
large cubic volume of comoving side length $L$ equipped with periodic boundary
conditions at each face. This is a construction commonly used in the study of
cosmological structure formation; the limit $L\rightarrow\infty$ can always be
taken as a final step. Furthermore, we divide the cubic volume into cells of
side length $d$ (i.e. we set $L=Nd$, $N>0$ an integer) since the initial
density perturbation (\ref{idc}) is periodic with comoving period $d$.

\subsubsection{Solution of the \fp \sch equation}

The solution of the \fp \sch equation in a static background cosmology is
discussed in detail in \ref{fpapp}. Rather than repeating the calculation for
the case of an expanding background, it suffices to note that we can simply
replace proper time $t$ by the linear growth factor $D$ in the
solution obtained in \ref{fpapp}. The solution of the \fp \sch equation (\ref{1dfptdse}) is then 

\begin{equation}
\label{fpsol}
\psi = \sum_{n}a_n\exp{\left[\frac{-\rmi(D-1)E_n^{(0)}}{\nu}\right]}\phi_n^{(0)},
\end{equation} 

\n where $E_n^{(0)}=\nu^2k_n^2/2$ and $k_n=2n\pi/d$ is a comoving
wavenumber. The (orthonormal) eigenfunctions $\phi_n^{(0)}=\phi_n^{(0)}(x)$ are of the form

\begin{equation}
\label{phin0}
\phi_n^{(0)}=\frac{1}{L^{3/2}}\exp{(\rmi k_n x)},
\end{equation}

\n and the expansion coefficients $a_n$ are determined from 

\begin{equation}
\label{an}
a_n=NL^2\int_{0}^{d}\overline{\phi_n^{(0)}}(x)\psi_{\rmi}(x)dx,
\end{equation}

\n where the over-line denotes complex conjugation and the initial
wavefunction $\psi_\rmi$ is given by (\ref{ipsi}). It follows from
(\ref{phin0}) that the expansion coefficients $a_n$ can be calculated simply
by taking the Fourier transform of the initial wavefunction. 

\subsection{Beyond the \fp approximation}
\label{1dst}

The \fp formalism provides an approximation to the full
Schr\"{o}dinger-Poisson system that is exactly solvable. The \fp approximation is motivated by the fact that the effective potential
$\mathcal{V}$ appearing in the Schr\"{o}dinger-Poisson system is zero to
first-order in both Eulerian and Lagrangian PT. Since the \fp approximation
is based on linear PT, a natural extension of the \fp  approximation could be
constructed simply by including higher-order  terms in PT. However, including
higher-order terms in PT implies $\mathcal{V}\neq 0$ in general
(e.g. \cite{cat1, cat2}), leading to a time-dependent external potential in the \sch equation (\ref{lDtdse}). The \sch equation with
a time-dependent potential is typically impossible to solve analytically, so
either a numerical or an approximate solution must be sought. 

In light of the above comments, we feel it is important to investigate whether
quantum-mechanical approximation methods can be used, in a cosmological
context, to obtain useful solutions of the \sch equation with a time-dependent
potential. We will focus on time-dependent perturbation theory (TDPT)
since this is one of the most widely used approximation schemes in
quantum mechanics. Our aim is to elucidate the properties of perturbative
solutions of the \sch equation with a time-dependent potential in a
cosmological setting. In particular, we wish to assess the effect of varying
the parameter $\nu$ where we again assume that $\nu$ is finite. It is not our intention here to explicitly formulate a
higher-order extension of the \fp approximation; we leave this for
future work. Instead we investigate perturbative solutions of the \sch
equation by considering a simple idealized scenario. We begin by rewriting the
Schr\"{o}dinger-Poisson system in a static background cosmology; the reason for this will become apparent. In a static universe the scale factor is constant $a=a_0=1$ (i.e. $H\equiv 0$) and the Friedmann equations imply that the universe is closed with a non-zero cosmological constant $\Lambda=4\pi G\rho_{\rm b,c}$. The Schr\"{o}dinger-Poisson system can then be written in the form

\begin{eqnarray}
\rmi\nu\frac{\partial\psi}{\partial t} = \left(-\frac{\nu^2}{2}\nabla_{\mathbf{x}}^{2} + \Phi\right)\psi,\label{tdse}\\
\nabla_{\mathbf{x}}^{2}\Phi - 4\pi G\rho_{\rm b,c}\left(|\psi|^{2}-1\right)=0,\label{wmpoiss}
\end{eqnarray}

\n where comoving coordinates $\mathbf{x}$ now coincide with physical coordinates $\mathbf{r}$. The potential $\Phi$ is the peculiar gravitational potential and 

\begin{equation}
\label{smad}
\psi=(1+\delta)^{1/2}\exp{\left(\frac{-\rmi\varphi}{\nu}\right)},
\end{equation}

\n where the peculiar velocity $\mathbf{v}$ is the gradient of the velocity
potential $\varphi=\varphi(\mathbf{x},t)$:
$\mathbf{v}=-\nabla_{\mathbf{x}}\varphi$. Inserting the Madelung transformation (\ref{smad}) into (\ref{tdse}) and (\ref{wmpoiss}) we obtain 

\begin{equation}
\label{sbern}
\frac{\partial\varphi}{\partial t}-\frac{1}{2}\left|\nabla_{\mathbf{x}}\varphi\right|^2-\Phi+\mathcal{P}=0,
\end{equation}

\n as well as the continuity equation (\ref{cty}) and the Poisson equation
(\ref{poiss}) in a static universe. The gradient of the modified Bernoulli
equation (\ref{sbern}) gives the Euler equation (\ref{eul}) in a static
background, but with an extra term corresponding to the gradient of the
quantum pressure $\mathcal{P}$. It follows from the fluid equations
(\ref{eul}), (\ref{cty}) and (\ref{poiss}) in a static background that, to
first-order in Eulerian PT, the peculiar gravitational potential evolves
according to $\Phi=D\Phi_\rmi$, where $D=\exp{[(t-t_\rmi)/\tau}]$ and
$\tau=1/\Lambda^{1/2}$ is the characteristic time-scale for the collapse of a
density fluctuation. To first-order in Lagrangian PT (i.e. the Zeldovich
approximation) it is straightforward to show from (\ref{phirel}) and (\ref{zbsol}) that, in a static universe, the gravitational potential $\Phi=\Phi(\mathbf{x},t)$ is given by 

\begin{equation}
\label{szapot}
\Phi=D\left[\Phi_\rmi-\frac{1}{2}(D-1)\left|\mathbf{v}_\rmi\right|^2\right],
\end{equation}

\n where $\Phi_\rmi=\Phi_\rmi(\mathbf{q})$ is the initial gravitational
potential, $\mathbf{v}_\rmi=\mathbf{v}_\rmi(\mathbf{q})$ is the initial
peculiar velocity field and the coordinates $\mathbf{x}$ and $\mathbf{q}$ are
related by the mapping (\ref{za}) with
$\mathbf{s}=\tau\mathbf{v}_\rmi$. Therefore, in a static background, we have a
non-zero time-dependent potential appearing in the \sch equation (\ref{tdse})
to {\it first-order}. This is in contrast to the expanding background case
where we must include higher-order terms in PT to obtain $\mathcal{V}\neq 0$
in the \sch equation (\ref{lDtdse}). Since our goal is to investigate
perturbative solutions of the \sch equation with a time-dependent potential,
it will clearly be simpler to work in a static background as we can then use a
first-order (rather than a cumbersome higher-order) expression for the
potential. This is why we have assumed a static background cosmology. 

To proceed with our investigation, we now apply the \wm approach to a specific
example of gravitational collapse. As before, we consider the evolution of a
plane-symmetric sinusoidal density perturbation $\delta_\rmi$ of the form
(\ref{idc}). The corresponding initial peculiar gravitational potential is
then found from the Poisson equation (\ref{wmpoiss}):

\begin{equation}
\label{slgpot}
\Phi_\rmi = -\left(\frac{d}{2\pi\tau}\right)^2\delta_\rmi,
\end{equation}

\n and the initial peculiar velocity field is given by

\begin{equation}
\label{viq}
v_\rmi^2=\left(\frac{d}{2\pi\tau}\right)^2\left(\delta_{\rm a}^2-\delta_\rmi^2\right),
\end{equation}

\n which follows from (\ref{slgpot}) by noting that, in a static universe, the
linear (Eulerian and Lagrangian) PT relation (\ref{phirel}) implies the
gravitational potential $\Phi$ and velocity potential $\varphi$ satisfy
$\varphi=\tau\Phi$. The full one-dimensional \sch equation we are aiming to
solve is then

\begin{equation}
\label{1dtdse} 
\rmi\nu\frac{\partial\psi}{\partial t}=\left(-\frac{\nu^2}{2}\frac{\partial^2}{\partial x^2}+\Phi\right)\psi,
\end{equation}

\n where we use the first-order Lagrangian PT expression (\ref{szapot})
for the gravitational potential $\Phi=\Phi(x,t)$ since this is exact in
one dimension (up until shell-crossing occurs). The initial gravitational
potential $\Phi_\rmi=\Phi_\rmi (q)$ and peculiar velocity $v_\rmi=v_\rmi(q)$
in (\ref{szapot}) are given by (\ref{slgpot}) and (\ref{viq}), respectively. The \sch equation (\ref{1dtdse}) must be solved subject to the initial condition 

\begin{equation}
\label{ismad}
\psi_\rmi=(1+\delta_\rmi)^{1/2}\exp{\left(\frac{-\rmi\varphi_\rmi}{\nu}\right)},
\end{equation}

\n where $\delta_\rmi$ is given by (\ref{idc}) and
$\varphi_\rmi=\tau\Phi_\rmi$. We again restrict our attention to a large cubic
volume of side length $L$ equipped with periodic boundary conditions at each
face. As before, the periodicity of the initial density perturbation suggests
it will be convenient to partition the cubic volume into cubic cells of side length $d$ by writing $L=Nd$, $N>0$ an integer.

\subsubsection{Perturbative solution of the \sch equation with a
  time-dependent potential}
\label{tdpot} 

The \sch equation (\ref{1dtdse}) cannot be solved analytically; we have used
TDPT to determine a second-order approximate solution instead. The details of the calculation are presented in \ref{tdapp}. We summarize the main results here for reference. The perturbative solution was found to be 

\begin{equation}
\label{tdwv}
\psi=\sum_{j=0}^2\psi^{(j)},
\end{equation}

\n where the zeroth-order term $\psi^{(0)}$ is simply the solution of the \fp
\sch equation ($\Phi\equiv 0$):

\begin{equation}
\label{td0}
\psi^{(0)}=\sum_n a_n\exp{\left[\frac{-\rmi (t-t_\rmi)E_{n}^{(0)}}{\nu}\right]}\phi_n^{(0)},
\end{equation}

\n with $E_n^{(0)}=\nu^2 k_n^2 /2$ and $k_n=2n\pi/d$. The eigenfunctions
$\phi_n^{(0)}$ are of the form (\ref{phin0}) and the expansion coefficients
$a_n$ are found from (\ref{an}) with the initial wavefunction $\psi_\rmi$ now given by (\ref{ismad}). The first-order term $\psi^{(1)}$ is 

\begin{eqnarray}
\label{td1}
\psi^{(1)} = -\rmi\frac{\delta_{\rm a}}{8\pi^2\gamma}\sum_{n}a_n\sum_{m}\mathcal{I}_{m,n}\phi_m^{(0)},
\end{eqnarray}

\n where $\gamma=\nu\tau/d^2$ is a dimensionless parameter, $\mathcal{I}_{m,n}=\mathcal{I}_{m,n}(t)$ is defined by 

\begin{equation}
\label{Imn}
\mathcal{I}_{m,n}=\frac{1}{\tau}\int_{t_\rmi}^{t}dt'\exp{\left[\frac{-\rmi (t-t')E_m^{(0)}}{\nu}\right]}\Phi_{m,n}(t')\exp{\left[\frac{-\rmi (t'-t_\rmi)E_{n}^{(0)}}{\nu}\right]},
\end{equation}

\n and the matrix elements $\Phi_{m,n}=\Phi_{m,n}(t)$ are given by

\begin{eqnarray}
\label{pmatel}
\fl\Phi_{m,n}=D\left\{\alpha\left[J_p(2p\alpha)+\frac{3}{2}\sum_{s=\pm 2}J_{p+s}(2p\alpha)\right]+\left(\frac{\alpha^2}{2}-1\right)\sum_{s=\pm 1}J_{p+s}(2p\alpha)\right.\nonumber\\
\left.-\frac{\alpha^2}{2}\sum_{s=\pm 3}J_{p+s}(2p\alpha)\right\},
\end{eqnarray}

\n where $p=m-n$ is an integer, $\alpha=\alpha(t)$ is defined by
$\alpha=\delta_{\rm a}(D-1)/2$ and the functions $J_l$ are Bessel functions of the first kind. The second-order term $\psi^{(2)}$ is given by

\begin{eqnarray}
\label{td2}
\psi^{(2)} = -\left(\frac{\delta_{\rm a}}{8\pi^2\gamma}\right)^2\sum_{n}a_n\sum_{m}\sum_{l}\mathcal{K}_{l,m,n}\phi_l^{(0)},
\end{eqnarray}

\n where $\mathcal{K}_{l,m,n}=\mathcal{K}_{l,m,n}(t)$ is defined by

\begin{eqnarray}
\label{Klmn}
\fl\mathcal{K}_{l,m,n}=\frac{1}{\tau^2}\int_{t_\rmi}^{t}dt'\int_{t_\rmi}^{t'}dt''\exp{\left[\frac{-\rmi (t-t')E_l^{(0)}}{\nu}\right]}\Phi_{l,m}(t')\exp{\left[\frac{-\rmi (t'-t'')E_{m}^{(0)}}{\nu}\right]}\nonumber\\
\times\Phi_{m,n}(t'')\exp{\left[\frac{-\rmi (t''-t_\rmi)E_{n}^{(0)}}{\nu}\right]}.
\end{eqnarray}

\n In order to complete the perturbation expansion (\ref{tdwv}) of the
wavefunction $\psi$, the integrals (\ref{Imn}) and (\ref{Klmn}) must be evaluated numerically. 

\section{Results and discussion}
\label{results}

We now analyze the gravitational collapse of the plane-symmetric
sinusoidal density perturbation (\ref{idc}) in (i) an expanding background,
using the \fp approximation, and (ii) a static background, using a
perturbative solution of the \sch equation with a time-dependent potential. In
both cases the CDM density field is obtained from the amplitude of the wavefunction via $\delta=|\psi|^2-1$ and compared with the linear growth law $\delta=D\delta_\rmi$ and the exact Zeldovich solution

\begin{equation}
\label{delc2}
\delta=\frac{(1+\delta_\rmi)}{\left[1-(D-1)\delta_\rmi\right]}-1,
\end{equation}

\n which follows from (\ref{delc}) and (\ref{deld}). The evolution of the
density fluctuation is followed up until particle trajectories cross which,
from (\ref{delc2}), occurs when the linear growth factor $D=D_{\rm
  sc}=1+1/\delta_{\rm a}$. Hereafter we will assume, for illustrative purposes, that the amplitude of the initial density perturbation is $\delta_{\rm a}=0.01$, so that $D_{\rm sc}=101$.

\subsection{Planar collapse using the \fp approximation}
\label{fpres}

The (exact) solution of the one-dimensional \fp \sch equation (\ref{1dfptdse})
is given by (\ref{fpsol}). Upon introducing the dimensionless
comoving coordinate $\bar{x}=x/d$, we find that the \fp solution depends
solely on the dimensionless parameter $\Gamma=\nu/d^2$, which we assume to be
finite (as opposed to $\Gamma\rightarrow 0$). Recall that, in the semi-classical limit $\Gamma\rightarrow 0$, the \fp
approximation coincides with the \za before shell-crossing occurs (see
\ref{path}). In one dimension this implies that the \fp approximation will be
exact in the limit $\Gamma\rightarrow 0$ (provided there is no
shell-crossing). This suggests that our numerical implementation of the \fp
approximation with finite $\Gamma$ can be optimized by using the smallest
possible value of $\Gamma$. In practice there is a lower
bound $\Gamma_{\rm c}$ on the value of $\Gamma$ arising from the fact that we
test the \fp approximation on a discrete grid. Sampling at the Nyquist rate
(the minimum possible sampling rate) requires that the phase change between
two neighbouring grid points must be less than or equal to $\pi$ radians. The phase of the initial wavefunction
is proportional to $1/\Gamma$ and so, as $\Gamma$ approaches zero, the
phase varies increasingly rapidly and the change in phase between two
neighbouring grid points can exceed $\pi$ radians. The phase is then
insufficiently sampled and aliasing effects cause our \fp method to break
down. To avoid this problem we must choose $\Gamma\geq\Gamma_{\rm c}$; we
expect the performance of the \fp approximation to be optimal for
$\Gamma=\Gamma_{\rm c}$. It is straightforward to show from (\ref{ipsi}) that
$\Gamma_{\rm c}=\delta_{\rm a}/2\pi^2 N_{\rm g}$, where $N_{\rm g}$ is the
number of grid points. We use $N_{\rm g}=512$ so that $\Gamma_{\rm c}=1\times
10^{-6}$. We now investigate how the \fp approximation, and in
particular the quantum pressure term, behaves as $\Gamma$ is decreased from a
large value towards $\Gamma=\Gamma_{\rm c}$. 

\begin{figure}[htbp]
\centering 
\epsfxsize=13cm 
\epsfbox{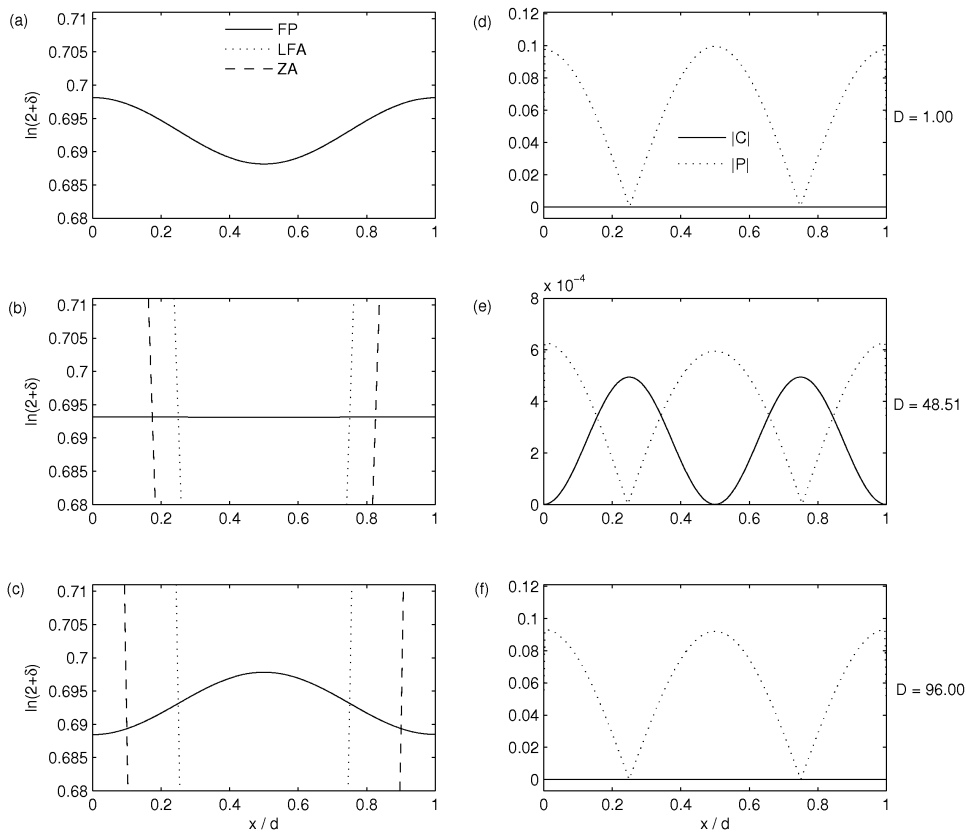}
\caption{Evolution of a plane-symmetric sinusoidal density perturbation in an
  expanding CDM-dominated universe. The amplitude of the initial density
  fluctuation is $\delta_{\rm a}=0.01$. The left-hand plots show the density
  fields obtained from the \fp approximation (FP), the linearized fluid
  approach (LFA) and the \za (ZA), at three different values of the linear
  growth factor. The parameter $\Gamma=1$ in the \fp approximation. The
  right-hand plots show the corresponding evolution of the magnitudes of the
  convective (C) and quantum pressure (P) terms, in units of $d^2$.} 
\label{fig1}
\end{figure}

 We begin by assuming $\Gamma=1$. The plots in the left-hand column of figure
 \ref{fig1} show the evolution of the \fp density field $\delta=|\psi|^2-1$ in
 this case. The density field is only plotted in the interval $0\leq x\leq d$ since, by construction, the density field is periodic with comoving period $d$. We can immediately see that the \fp density field simply oscillates about the mean value $\langle\delta\rangle=0$ and there is no net growth of the initial
 density perturbation. We find that this oscillatory behaviour also persists
 for any $\Gamma>1$. This effect is caused by the quantum pressure term. Recall that the quantum pressure $\mathcal{P}$ enters the \fp formalism
via the modified Bernoulli equation (\ref{bern3}). Since
$\mathcal{P}\propto\nu^2\propto\Gamma^2$, the parameter $\Gamma$ controls the
size of the quantum pressure term relative to the convective term
$\mathcal{C}=\mathcal{C}(\mathbf{x},D)$, defined by
$\mathcal{C}=-\left|\nabla_{\mathbf{x}}\phi\right|^2/2$. The plots in the
right-hand column of figure \ref{fig1} compare the magnitudes of the quantum
pressure and convective terms as a function of the linear growth factor. The
quantum pressure term is the dominant term appearing in (\ref{bern3}) at all
times (except when it momentarily becomes zero as the
density field passes through $\delta\equiv 0$). Therefore, the growth of the
density perturbation appears to be inhibited by the
large quantum pressure term. This is somewhat reminiscent of the \ada where
 the growth of fluctuations is suppressed on scales $\lesssim\mu^{1/2}$, leading to the
large-scale structure distribution appearing `washed out' for large $\mu$
\cite{wein}. However, even if the viscosity term is large, the density field in
the adhesion model does not display oscillatory behaviour; this is a
peculiarity of the quantum pressure term. It is interesting to note that gas
pressure in a baryonic fluid causes a qualitatively similar effect
(e.g. \cite{peeb}). If we were considering a baryonic fluid (rather than a
pressureless fluid of CDM) then it would be necessary to account for the
effects of gas pressure by including a classical pressure term in the Euler
equation. Applying first-order Eulerian PT to the fluid equations then yields
a characteristic length scale (the {\it Jean's length}); any Fourier modes of
the density field with a wavelength smaller than the Jean's length undergo
damped oscillation rather than growth. However, we must be careful not to
directly compare the effects of the quantum pressure term to those of a classical pressure term since the origin and form of the two terms is very different.

\begin{figure}[htbp]
\centering 
\epsfxsize=13cm
\epsfbox{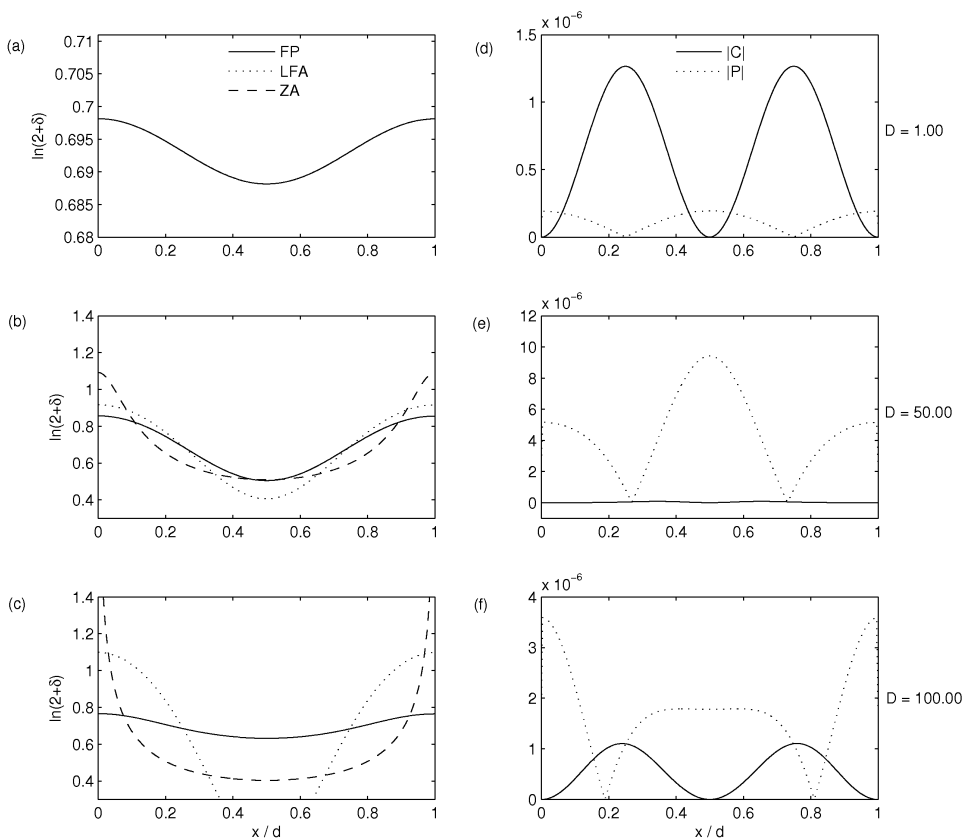}
\caption{Evolution of a plane-symmetric sinusoidal density perturbation in an
  expanding CDM-dominated universe. The layout of the plots is as in figure
  \ref{fig1}; the only difference is that the parameter $\Gamma=1.4\times 10^{-3}$ in the \fp approximation.} 
\label{fig2}
\end{figure}

The quantum pressure $\mathcal{P}\propto\Gamma^2$ should become less
significant as $\Gamma$ is decreased. The left-hand column of figure \ref{fig2} displays the
\fp density field as a function of the linear growth factor when
$\Gamma=1.4\times 10^{-3}$. Figure \ref{fig2}(a) and figure \ref{fig2}(b) show
that the \fp density field evolves as expected up to $D=D_{\rm sc}/2$, in the
sense that matter flows towards initially over-dense regions. However, the growth rate of
the over-densities is considerably less than that predicted by the exact
Zeldovich solution. For $D>D_{\rm sc}/2$, the over-densities in the \fp
density field cease to grow and actually begin to decay, leading to a large
discrepancy between the \fp and \za density fields at $D\approx D_{\rm sc}$;
see figure \ref{fig2}(c). This behaviour is counter-intuitive; over-densities
should continue to grow up until shell-crossing occurs as in first-order
Eulerian and Lagrangian PT. The quantum pressure term $\mathcal{P}$ again
provides the explanation. The right-hand column of figure \ref{fig2} shows the
evolution of $|\mathcal{C}|$ and $|\mathcal{P}|$. We
can see that the convective term initially dominates the quantum pressure term
in the Bernoulli equation (\ref{bern3}). This remains the case at early times,
corresponding to when the \fp approximation agrees well with the linearized fluid approach and the Zeldovich
approximation. However, the quantum pressure quickly grows until, at $D\approx
D_{\rm sc}/2$, it is much larger than the convective term. The large quantum
pressure term suppresses the gravitational collapse of the density perturbation, causing the over-densities to cease growing and subsequently decay. 

\begin{figure}[htbp]
\centering 
\epsfxsize=13cm
\epsfbox{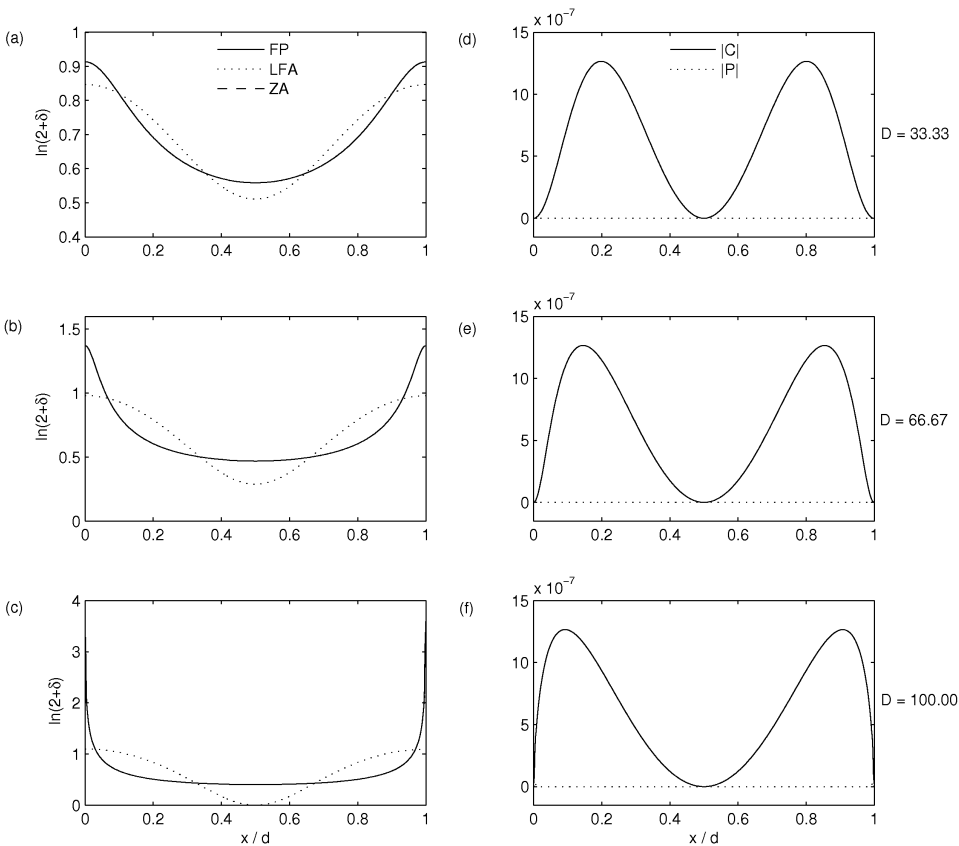}
\caption{Evolution of a plane-symmetric sinusoidal density perturbation in an
  expanding CDM-dominated universe. The layout of the plots is as in figure
  \ref{fig1}; the only difference is that the parameter $\Gamma=\Gamma_{\rm c}=1\times 10^{-6}$ in the \fp approximation.} 
\label{fig3}
\end{figure}

The effect of the quantum pressure term is indeed less pronounced in the 
$\Gamma=1.4\times 10^{-3}$ case than in the $\Gamma=1$ case. However, at
late times it causes matter to flow away from over-dense regions in an
unrealistic manner. We find that, when $\Gamma$ is reduced to $\Gamma\approx
1\times 10^{-3}$, the quantum pressure never becomes large enough to cause the
decay of over-densities and the initial density perturbation grows up until
shell-crossing. However, the collapse of the fluctuation is still highly
suppressed by the quantum pressure term in high-density regions and the \fp
approximation thus provides a poor match to the \za in such regions. As
$\Gamma\rightarrow\Gamma_{\rm c}$, the suppression effect of the quantum
pressure term diminishes and the agreement between the \fp and Zeldovich
approximations improves. The left-hand column of figure \ref{fig3} shows the
\fp evolution of the initial density perturbation in the optimal case
$\Gamma=\Gamma_{\rm c}=1\times 10^{-6}$. At all values of the linear growth factor the \fp approximation now provides an excellent match
to the exact Zeldovich solution, even in over-dense regions. However, unlike
the Zeldovich approximation, the \fp approximation leads to a density field
that remains well behaved at shell-crossing. The right-hand column of figure
\ref{fig3} shows the corresponding evolution of $|\mathcal{C}|$ and
$|\mathcal{P}|$. It is immediately apparent that the convective term
completely dominates the quantum pressure term at all values of the linear
growth factor. Although the quantum pressure term is small, it is non-zero
(because we are using a finite $\Gamma$) and thus will have some effect on the
collapse of the density perturbation. However, this effect is negligibly small for $\Gamma=1\times 10^{-6}$. 

\subsection{Planar collapse using the \sch equation with a time-dependent potential} 
\label{1dtdp}

In a static background the linear growth factor is given by
$D=\exp{[(t-t_\rmi)/\tau]}$. Upon introducing a dimensionless time coordinate
$\bar{t}=(t-t_\rmi)/\tau$ we find that, for $\delta_{\rm a}=0.01$,
shell-crossing occurs when $\bar{t}=\bar{t}_{\rm sc}=\ln{(D_{\rm sc})}\approx
4.6$. We now use the perturbative solution (\ref{tdwv}) of the one-dimensional
\sch equation with a time-dependent potential (\ref{1dtdse}) to follow the
gravitational collapse of the initial density perturbation (\ref{idc}) up to
$\bar{t}_{\rm sc}$. Observe that the perturbation expansion of the
wavefunction depends on one dimensionless parameter: $\gamma=\nu\tau/d^2$,
which we assume to be finite. As in the previous section, phase-aliasing
effects impose a lower bound $\gamma_{\rm c}$ on the value of $\gamma$
in practice. It follows from (\ref{ismad}) that again $\gamma_{\rm
  c}=\delta_{\rm a}/2\pi^2 N_{\rm g}=1\times 10^{-6}$. Based on the results of
our test of the \fp approximation, it is reasonable to suppose that the
performance of the \wm approach will be optimized for
$\gamma=\gamma_{\rm c}$. Since the gravitational potential (\ref{szapot})
appearing in (\ref{1dtdse}) is exact in one-dimension, we would expect the \wm density field to agree well with that of the exact Zeldovich
solution in this optimal case. Unfortunately, as we shall see, the
situation is more complex when dealing with perturbative solutions of the \sch
equation. We now examine the behaviour of the TDPT solution of the \sch
equation as $\gamma$ is reduced from a large value towards $\gamma=\gamma_{\rm c}$.

\begin{figure}[htbp]
\centering 
\epsfxsize=13cm
\epsfbox{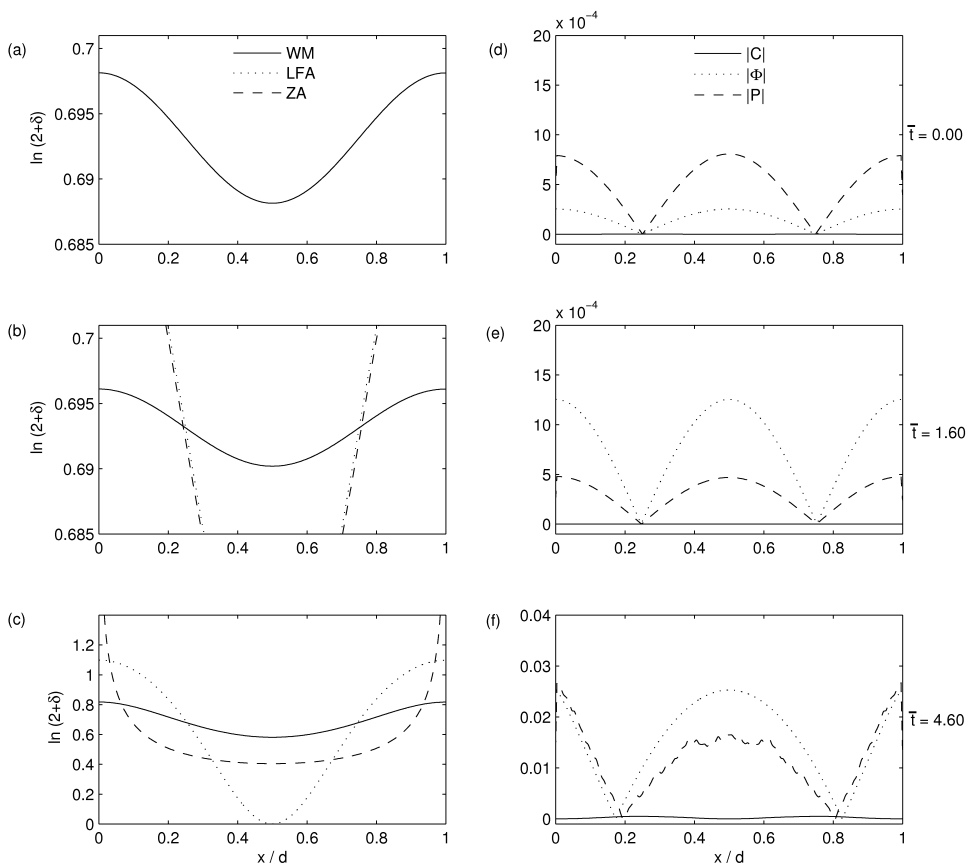}
\caption{Evolution of a plane-symmetric sinusoidal density perturbation in a
  static CDM-dominated universe. The amplitude of the initial density
  perturbation is $\delta_{\rm a}=0.01$. The left-hand plots show the density
  fields obtained from the \wm approach with a time-dependent
  potential (WM), the linearized fluid approach (LFA) and the \za (ZA). The parameter $\gamma=0.1$ in the \wm approximation. The right-hand plots show the corresponding evolution of the magnitudes of the convective (C), gravitational potential ($\Phi$) and quantum pressure (P) terms, in units of $d^2/\tau^2$.} 
\label{fig4}
\end{figure}

As a starting point we choose $\gamma=1$. In this case we find that the \wm
density field $\delta=|\psi|^2-1$ oscillates rapidly about
$\langle\delta\rangle=0$ and there is no overall growth of the initial density
fluctuation (cf. figure 1). In fact, the \wm approach leads to an oscillatory
density field for any $\gamma>1$. As in the previous section, this is due to
the fact that the quantum pressure term is the dominant term in the Bernoulli
equation (\ref{sbern}) for all times up to shell-crossing. The quantum pressure
$\mathcal{P}\propto\nu^2\propto\gamma^2$ and thus should become less
influential as $\gamma$ is decreased. The evolution of the \wm density field
for $\gamma=0.1$ is shown in the left-hand column of figure \ref{fig4}. By
comparing figure \ref{fig4}(a) with figure \ref{fig4}(b) we see that the
initial over-densities begin to decay rather than grow; this is
unphysical. However, at $\bar{t}\approx \bar{t}_{\rm sc}/3$, the decay slows
and halts and matter begins to move towards over-dense regions in the usual fashion up to
$\bar{t}\approx\bar{t}_{\rm sc}$. Since over-dense regions only begin to
collapse after $\bar{t}\approx\bar{t}_{\rm sc}/3$,
the final \wm density field is considerably more homogeneous than the density
fields obtained from the linearized fluid approach and the Zeldovich
approximation. The reason for this behaviour becomes apparent if we examine
the relative sizes of the convective, gravitational potential and quantum
pressure terms in (\ref{sbern}). The plots in the right-hand column of figure
\ref{fig4} compare the time-variation of $|\mathcal{C}|$, $|\Phi|$ and
$|\mathcal{P}|$. At $\bar{t}=0$ the quantum pressure is the dominant term and
subsequently causes the initial over-densities to decay. However, in a static
background, the gravitational potential (\ref{szapot}) grows rapidly with time
and, at $\bar{t}\approx \bar{t}_{\rm sc}/3$, dominates over the quantum
pressure and convective terms. This corresponds to the time when the
over-densities cease to decay and the gravitational collapse of the density
fluctuation begins to proceed as expected. The gravitational potential then remains the dominant term up until shell-crossing occurs. 

\begin{figure}[htbp]
\centering 
\epsfxsize=13cm
\epsfbox{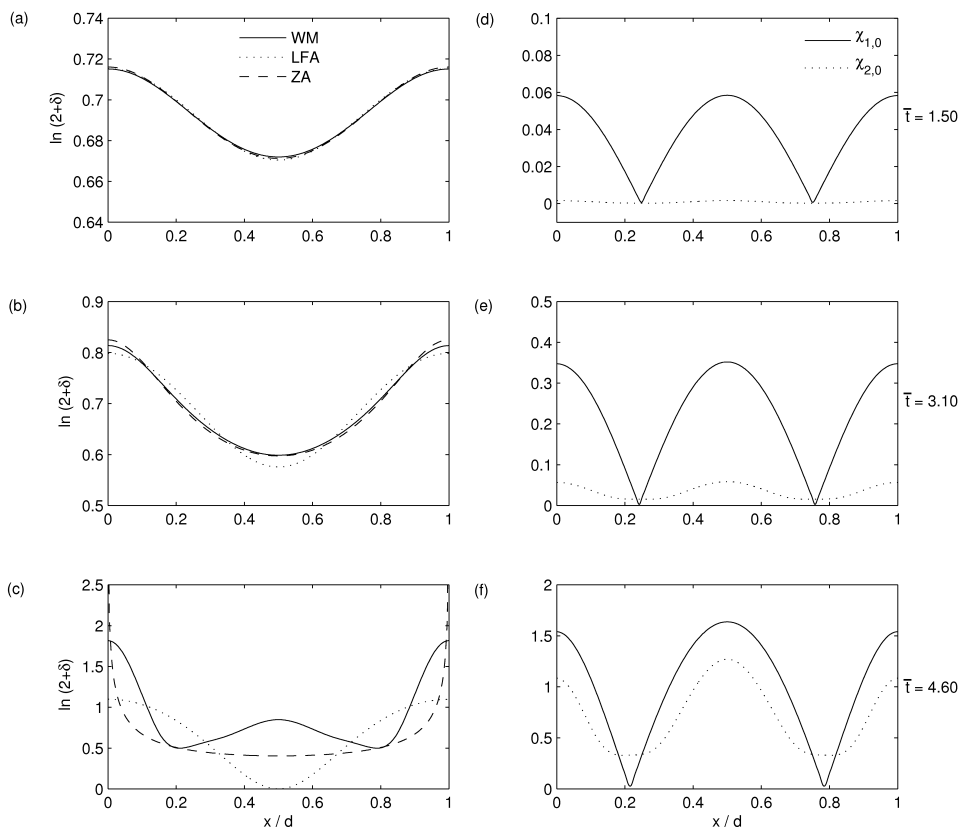}
\caption{Evolution of a plane-symmetric sinusoidal density perturbation in a
  static CDM-dominated universe. The left-hand plots are as in figure
  \ref{fig4}, except that the parameter $\gamma=0.015$ in the \wm
  approximation. The right-hand plots show the corresponding evolution of the ratios $\chi_{1,0}$ and $\chi_{2,0}$ defined in the text.} 
\label{fig5}
\end{figure}

We find that, as the value of $\gamma$ is reduced further, the effect of the
quantum pressure term indeed lessens. However, for $\gamma<0.02$, the \wm approach exhibits rather unusual behaviour. For example, consider $\gamma=0.015$ (note $\gamma\gg\gamma_{\rm c}$); figure \ref{fig5} illustrates
the behaviour of the \wm approximation in this case. It is evident
from figure \ref{fig5}(a) and figure \ref{fig5}(b) that the \wm approximation
now provides a close match to the linearized fluid approach and the exact
Zeldovich solution up to $\bar{t}\approx 2\bar{t}_{\rm sc}/3$. However, at
times close to shell-crossing, an over-density appears at $x=d/2$ (i.e. where
the gravitational potential is at a {\it maximum}) in the \wm density
field. This is clearly not realistic. To explain this, recall that the
second-order perturbative solution of the \sch equation (\ref{1dtdse}) is 

\begin{equation}
\label{pertexp}
\psi=\sum_{j=0}^{2}\psi^{(j)},
\end{equation} 

\n where $\psi^{(0)}$, $\psi^{(1)}$ and $\psi^{(2)}$ are given by (\ref{td0}),
(\ref{td1}) and (\ref{td2}), respectively. In order for the perturbation
expansion (\ref{pertexp}) to be valid we require the first and second-order
terms to be much smaller than the zeroth-order term. However,
$\psi^{(1)}\propto 1/\gamma$ and $\psi^{(2)}\propto 1/\gamma^2$ so, for small
values of $\gamma$, we anticipate that these terms will be large compared to
$\psi^{(0)}$. To check this we define a ratio 
 
\begin{equation}
\label{chi2}
\chi_{j,0}=\frac{|\psi^{(j)}|}{|\psi^{(0)}|},
\end{equation}

\n with $j=1,2$. We then require $\chi_{1,0}\ll 1$ and $\chi_{2,0}\ll 1$ for our perturbative solution to hold. The ratios $\chi_{1,0}$ and $\chi_{2,0}$ are shown as functions of time in the right-hand column of figure \ref{fig5}. At early times $\chi_{1,0}$ and $\chi_{2,0}$ are both small as they should
be. However, they both grow with time until, at $\bar{t}\approx \bar{t}_{\rm
  sc}$, the first and second-order terms in the perturbation expansion of the
wavefunction are comparable to the zeroth-order term. A general feature is
that the first-order term initially grows at a faster rate than the
second-order term but, as the time of shell-crossing approaches, the
second-order term begins to grow very rapidly and is soon comparable to the
first-order term. The false over-density at $x=d/2$ then develops, suggesting
that this is due to the onset of non-perturbative behaviour in the system. We
find that, as $\gamma\rightarrow\gamma_{\rm c}$, the perturbative solution of
the \sch equation breaks down at progressively earlier times. 

The development of non-perturbative behaviour is somewhat reminiscent of the
onset of turbulence in a classical fluid. It is interesting to speculate that
small values of $\gamma$ could be linked to some form of turbulent behaviour
in our hydrodynamical description of quantum mechanics, especially when we
realise that $\nu$ has dimensions of viscosity and that the dimensionless
parameter $1/\gamma$ is similar to the Reynold's number appearing in classical
fluid mechanics. Turbulence in classical fluids is associated with the
formation of vortices which obviously does not occur in one-dimensional
systems. Nevertheless the analogy is compelling. A possible explanation is
that very small values of $\gamma$ involve large angular frequencies in the
phase of the wavefunction (which is proportional to $1/\gamma$). To regain
classical behaviour in the limit $\gamma\rightarrow 0$ the waves from which
$\psi$ is constructed must cancel exactly. Since the limit $\gamma\rightarrow
0$ involves waves of infinitely small wavelength, any finite perturbative
calculation will find convergence impossible. A related but converse
phenomenon arises in Eulerian fluids (i.e. fluids without intrinsic viscosity)
when they are described in terms of a truncated spectrum \cite{cich}.

\begin{figure}[htbp]
\centering 
\epsfxsize=13cm
\epsfbox{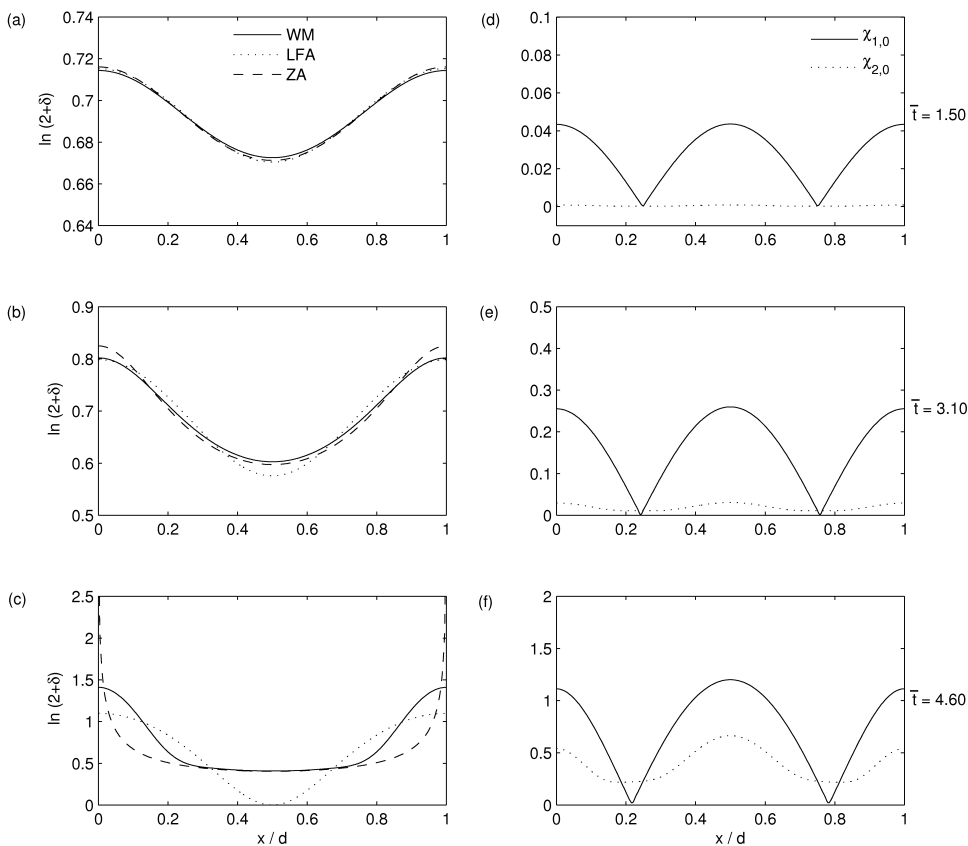}
\caption{Evolution of a plane-symmetric sinusoidal density perturbation in a
  static CDM-dominated universe. The layout of the plots is as in figure \ref{fig5}; the only difference is that the parameter $\gamma=0.02$ in the \wm approximation.} 
\label{fig6}
\end{figure}

It is clear from the above discussion that there is a conflict to be faced
when choosing an appropriate value of $\gamma$. On the one hand, we wish to
set $\gamma=\gamma_{\rm c}$ to ensure that the quantum pressure term is
minimized. On the other hand, the accuracy of our perturbative solution of the
\sch equation improves as $\gamma$ is {\it increased}. The best one can do in
this situation is to compromise by selecting the smallest possible value of
$\gamma$ for which the perturbative solution of the \sch equation remains
physically reasonable over the time-scales of interest. We denote this
`optimal' value by $\gamma_{\rm o}$. In the situation under consideration
here, we find that $\gamma_{\rm o}=0.02$; the left-hand column
of figure \ref{fig6} shows the behaviour of the \wm density field in this
case. We can see from figure \ref{fig6}(a) and figure \ref{fig6}(b) that the
\wm approach is again in good agreement with the linearized fluid approach and
the exact Zeldovich solution for times up to $\bar{t}\approx 2\bar{t}_{\rm
  sc}/3$. Figure \ref{fig6}(c) shows that, at $\bar{t}\approx\bar{t}_{\rm
  sc}$, the \wm density field is well-behaved (in the sense that there is no
unphysical over-density at $x=d/2$) and agrees extremely
well with the \za in the vicinity of $x=d/2$. The \wm approach also leads to
the formation of over-densities with $\delta\approx 2.2$ (i.e. in the
quasi-linear regime). However, there is still a large
discrepancy between the \wm and \za density fields in high-density
regions. This is partly attributable to the fact that, although dominated by
the gravitational potential term, the quantum pressure term is still
non-negligible for $\gamma=0.02$ and thus hinders the collapse of the density
fluctuation. In addition, as shown by the plots in the right-hand column of
figure \ref{fig6}, the ratios $\chi_{1,0}$ and $\chi_{2,0}$ grow with time,
eventually becoming comparable to unity at $\bar{t}\approx\bar{t}_{\rm
  sc}$. Consequently, our perturbative solution (\ref{tdwv}) of the \sch
equation is on the verge of failing completely at times close to
shell-crossing. 

\section{Conclusion}
\label{conc}

In this paper we have proposed a new approximation method suitable for evolving
cosmological density perturbations into the quasi-linear regime. This new
method is based upon the \fp \sch equation and is thus called the \fp
approximation. We have seen that the \fp approximation provides an alternative
to the well-known adhesion model in which the viscosity term
$\mu\nabla_{\mathbf{x}}^2\phi$ is replaced by the quantum pressure
term $\mathcal{P}$. The quantum pressure acts as a regularizing term that
prevents the generation of density singularities when particle trajectories
cross. Therefore, in contrast to the Zeldovich approximation, the \fp method
does not fail catastrophically when shell-crossing occurs. The quantum
pressure term is controlled by a free parameter $\nu$ which plays a similar
role to the viscosity parameter $\mu$. We have shown that, in the
semi-classical limit $\nu\rightarrow 0$, the \fp approximation reduces to the
Zeldovich approximation, provided there is no shell-crossing. The quantum
pressure term will then only have an effect in multi-stream regions.

We have performed a simple test of the \fp approximation by considering the
gravitational collapse of a plane-symmetric sinusoidal density
perturbation. The behaviour of the \fp approximation has been examined for
several different finite values of the parameter $\nu$ (or, equivalently,
$\Gamma$ in the notation of section \ref{fpres}) in an attempt to elucidate
the role of the quantum pressure term. We have found that, contrary to the
$\nu\rightarrow 0$ case, the quantum pressure affects the \fp approximation
before shell-crossing occurs if $\nu$ is finite. In particular, the quantum
pressure acts to suppress the gravitational collapse of density fluctuations;
the degree of suppression depends on the actual value of $\nu$. For large
values of $\nu$, the quantum pressure term completely dominates over the convective term in the modified Bernoulli equation (\ref{bern3}) and the
\fp density field then simply oscillates about $\langle\delta\rangle =0$. As
the value of $\nu$ is decreased the effect of
the quantum pressure term diminishes. When the value of $\nu$ is set to the
smallest possible value $\nu_{\rm c}$ allowed by our numerical implementation
of the \fp method, we find that the \fp approximation provides an excellent match to the exact Zeldovich solution of the equations
of motion, up until shell-crossing occurs. In this case the quantum pressure
term is still non-zero, but is negligibly small in comparison with the
convective term and thus has little effect before shell-crossing.

The success of the \fp approximation in the simple example presented here
suggests that this method promises to be a useful analytical tool for
modelling the formation of large-scale structure in the universe. In a
subsequent paper we will exhaustively test the \fp approximation in a more
cosmologically relevant scenario by appealing to a full \nb simulation
\cite{short}.

The \fp approximation is motivated by the fact that, to first-order in
Eulerian and Lagrangian PT, the effective potential $\mathcal{V}$ appearing in
the full Schr\"{o}dinger-Poisson system is zero. In future work it may prove
desirable to go beyond the \fp approximation by including
higher-order terms in PT. We anticipate that any such extension of the \fp
approximation will, in general, lead to a \sch equation with a time-dependent
potential that cannot be solved exactly. Instead, either numerical or
approximation methods must be deployed. In this work we have examined
whether TDPT is a suitable technique for constructing approximate solutions of
the \sch equation in a cosmological setting. We have performed our
investigation by appealing to a simple model where the universe is assumed to
be static. In this case the gravitational potential $\Phi$ appearing in the
Schr\"{o}dinger-Poisson system is non-zero to first-order in both Eulerian and
Lagrangian PT. We use first-order Lagrangian PT (which is exact in
one-dimension) to calculate $\Phi$ in the \sch equation for the case of a
plane-symmetric sinusoidal density perturbation. A perturbative solution of
the resulting \sch equation was then found by applying second-order TDPT. To
elucidate the properties of this perturbative solution we have studied the
behaviour of the \wm approximation as the (finite) parameter $\nu$ (or
$\gamma$ in the notation of section \ref{1dtdp}) is varied. For large values
of $\nu$ the quantum pressure term $\mathcal{P}$ is the dominant term in the
modified Bernoulli equation (\ref{sbern}) and again acts to inhibit the
gravitational collapse of the density fluctuation. On the other hand, we have
seen that the perturbation expansion of the wavefunction breaks down for small
values of $\nu$, leading to the formation of an unphysical over-density where
the gravitational potential is at a maximum. Therefore, in order to optimize the
performance of our perturbative \wm approach, we must be satisfied with
choosing the smallest possible value of $\nu$ for which the TDPT solution of
the \sch equation remains physically sensible over the time-scales of
interest. In this optimal case we find that the \wm approach is capable of
evolving the density perturbation into the quasi-linear regime, but provides a
poor match to the exact solution in over-dense regions.

In summary, we have shown that TDPT is a viable means of calculating
approximate solutions of the \sch equation in a cosmological context, provided
extreme care is taken in choosing $\nu$. Although we have only considered a
simple example, we believe that the issues we have highlighted
regarding perturbative solutions of the \sch equation will remain pertinent in
more general situations. Thus our work promises to be useful in the future
when constructing more sophisticated alternatives to the \fp approximation.

\ack

C J Short would like to thank PPARC for the award of a studentship that made
this work possible. We also thank the anonymous referee for helpful comments.

\appendix

\section{Path integral solution of the \fp \sch equation in the semi-classical limit}
\label{path}

The solution $\psi=\psi(\mathbf{x},D)$ of the \fp \sch equation

\begin{equation}
\rmi\nu\frac{\partial\psi}{\partial D} = -\frac{\nu^2}{2}\nabla_{\mathbf{x}}^{2}\psi
\end{equation}

\n can be written in the form

\begin{equation}
\label{psik}
\psi = \int U(\mathbf{x},D;\mathbf{q},1)\psi_{\rm i}(\mathbf{q}) d^3\mathbf{q},
\end{equation}

\n where $\psi_{\rm i}=\psi_{\rm i}(\mathbf{q})$ is some initial wavefunction and $U=U(\mathbf{x},D;\mathbf{q},1)$ is the {\it \fp propagator}. In the path integral formulation of quantum mechanics, the propagator involves a sum over all possible space-time paths connecting the points $(\mathbf{q},1)$ and $(\mathbf{x},D)$. The path integral
  expression for the \fp propagator is a standard result, see \cite{fey} for example. In the case at hand the \fp propagator is given by

\begin{equation}
\label{kfp}
U=\frac{1}{[2\pi\rmi\nu(D-1)]^{3/2}}\exp{\left[\frac{\rmi|\mathbf{x}-\mathbf{q}|^2}{2\nu(D-1)}\right]},
\end{equation}

\n and so the wavefunction (\ref{psik}) becomes

\begin{equation}
\label{psik2}
\psi =\frac{1}{[2\pi\rmi\nu(D-1)]^{3/2}}\int\exp{\left[\frac{\rmi|\mathbf{x}-\mathbf{q}|^2}{2\nu(D-1)}\right]}\psi_{\rm i}(\mathbf{q}) d^3\mathbf{q}.
\end{equation}

\n Inserting the Madelung form 

\begin{equation}
\psi_{\rm i}=(1+\delta_{\rm i})^{1/2}\exp{\left(\frac{-\rmi\phi_{\rm i}}{\nu}\right)}
\end{equation}

\n of the initial wavefunction into (\ref{psik2}) then yields

\begin{equation}
\label{psik3}
\fl\psi = \frac{1}{[2\pi\rmi\nu(D-1)]^{3/2}}\int\left[1+\delta_{\rm i}(\mathbf{q})\right]^{1/2}\exp{\left(\frac{\rmi}{\nu}\left[\frac{|\mathbf{x}-\mathbf{q}|^2}{2(D-1)}-\phi_{\rm i}(\mathbf{q})\right]\right)} d^3\mathbf{q}.
\end{equation}

\n Recall that we are interested in the solution of the \fp \sch equation in
the semi-classical limit $\nu\rightarrow 0$. The exponent appearing in
(\ref{psik3}) is complex and so the integrand will be a rapidly oscillating
function of $\mathbf{q}$ in the limit $\nu\rightarrow 0$. The dominant contribution to the integral in
(\ref{psik3}) will then be from points where the phase varies least rapidly
with $\mathbf{q}$, i.e. at stationary points. This method of evaluating the
integral in (\ref{psik3}) is referred to as the {\it method of stationary
  phase}. Before proceeding, we first rewrite the exponent in (\ref{psik3}) by introducing a function $S=S(\mathbf{x},D;\mathbf{q},1)$ of the form

\begin{equation}
\label{S}
S=\frac{|\mathbf{x}-\mathbf{q}|^2}{2(D-1)}-\phi_{\rm i}.
\end{equation}

\n Hereafter we will write $S=S(\mathbf{q})$ for brevity since the integral in
(\ref{psik3}) is over $\mathbf{q}$ only. At stationary points
$\bar{\mathbf{q}}$ of the phase we then have $S_{,m}(\bar{\mathbf{q}})=0$,
where $S_{,m}$ denotes $\partial S/\partial q_m$, leading to

\begin{equation}
\label{sp}
\mathbf{x}=\bar{\mathbf{q}}+(D-1)\mathbf{s}(\bar{\mathbf{q}}),
\end{equation}

\n with the time-independent vector field $\mathbf{s}=\mathbf{s}(\mathbf{q})$
defined by $\mathbf{s}=-\nabla_{\mathbf{q}}\phi_{\rm i}$. In what follows
we assume that, for a given value $D$ of the linear growth factor, there is
one and only one stationary point $\bar{\mathbf{q}}$ satisfying (\ref{sp})
for each $\mathbf{x}$. This is tantamount to assuming the absence of
shell-crossing. We now Taylor expand the function $S$ about the stationary point $\bar{\mathbf{q}}$:

\begin{equation}
\label{tay}
S(\mathbf{q})\approx S(\bar{\mathbf{q}})+\frac{1}{2}\sum_{m=1}^{3}\sum_{n=1}^{3}(q_m - \bar{q}_m)S_{,mn}(\bar{\mathbf{q}})(q_n -\bar{q}_n),
\end{equation} 

\n where $S_{,mn}$ denotes $\partial^2 S/\partial q_m\partial q_n$ and we have
used the fact that $S_{,m}(\bar{\mathbf{q}})=0$. Inserting (\ref{tay}) into
(\ref{psik3}) and changing variables to $\mathbf{p}=\mathbf{q}-\bar{\mathbf{q}}$ we obtain

\begin{equation}
\label{psik4}
\psi = \frac{\left[1+\delta_{\rm i}(\bar{\mathbf{q}})\right]^{1/2}}{[2\pi\rmi\nu(D-1)]^{3/2}}\exp{\left[\frac{\rmi}{\nu}S(\bar{\mathbf{q}})\right]}\int\exp{\left[-\frac{1}{2}\mathbf{p}^{T}\mathbf{M}(\bar{\mathbf{q}})\mathbf{p}\right]} d^3\mathbf{p},
\end{equation}

\n where the superscript $T$ denotes a transpose and
$\mathbf{M}(\bar{\mathbf{q}})$ is a complex symmetric matrix whose elements
are $M_{mn}(\bar{\mathbf{q}})=-\rmi S_{,mn}(\bar{\mathbf{q}})/\nu$. The
three-dimensional Gaussian integral in (\ref{psik4}) can be evaluated using standard techniques:

\begin{equation}
\label{gauss}
\int\exp{\left[-\frac{1}{2}\mathbf{p}^{T}\mathbf{M}(\bar{\mathbf{q}})\mathbf{p}\right]} d^3\mathbf{p}=\left[\frac{(2\pi)^{3}}{\mathcal{M}(\bar{\mathbf{q}})}\right]^{1/2},
\end{equation}

\n where $\mathcal{M}(\bar{\mathbf{q}})$ is the determinant of
  $\mathbf{M}(\bar{\mathbf{q}})$. In order for (\ref{gauss}) to hold we
  require $\mathcal{M}(\bar{\mathbf{q}})\neq 0$; this condition is satisfied
  before shell-crossing occurs. To see this, first note that (\ref{S}) implies
  $S_{,mn}(\bar{\mathbf{q}})=J_{mn}(\bar{\mathbf{q}})/(D-1)$ where 
  $J_{mn}(\bar{\mathbf{q}})=\delta_{mn}+(D-1)s_{m,n}(\bar{\mathbf{q}})$ and
  $\delta_{mn}$ is the Kronecker delta. The real symmetric matrix
  $\mathbf{J}(\bar{\mathbf{q}})$ with elements $J_{mn}(\bar{\mathbf{q}})$ is
  then simply the Jacobian matrix of the \za evaluated at the stationary point
  $\bar{\mathbf{q}}$. It follows that

\begin{equation}
\label{MJ}
\mathcal{M}(\bar{\mathbf{q}})=\frac{\rmi}{\left[\nu (D-1)\right]^3}\mathcal{J}(\bar{\mathbf{q}}),
\end{equation}

\n where $\mathcal{J}(\bar{\mathbf{q}})$ is the determinant of the Jacobian
matrix $\mathbf{J}(\bar{\mathbf{q}})$. Since $\mathcal{J}(\bar{\mathbf{q}})$
is non-zero up until shell-crossing occurs, then so is
$\mathcal{M}(\bar{\mathbf{q}})$. We can then substitute (\ref{MJ}) and
(\ref{gauss}) into (\ref{psik4}) to obtain

\begin{equation}
\label{psik5}
\psi = \left[\frac{1+\delta_{\rm i}(\bar{\mathbf{q}})}{\mathcal{J}(\bar{\mathbf{q}})}\right]^{1/2}\exp{\left[\frac{\rmi}{\nu}S(\bar{\mathbf{q}})\right]}.
\end{equation}
  
\n Comparing (\ref{psik5}) with the Madelung transformation we see that, in
the limit $\nu\rightarrow 0$, the CDM density field $\delta=\delta(\mathbf{x},D)$ in the \fp approximation is given by

\begin{equation}
\label{fpdn}
\delta =\frac{\left[1+\delta_{\rm i}(\bar{\mathbf{q}})\right]}{\mathcal{J}(\bar{\mathbf{q}})}-1,
\end{equation}

\n and the velocity potential $\phi=\phi(\mathbf{x},D)$ follows from (\ref{S}):

\begin{equation}
\label{fppn}
\phi=\phi_\rmi(\bar{\mathbf{q}})-\frac{1}{2}(D-1)\left|\mathbf{s}(\bar{\mathbf{q}})\right|^2,
\end{equation}

\n where $\mathbf{x}$ and $\bar{\mathbf{q}}$ satisfy (\ref{sp}). The
expressions (\ref{fpdn}) and (\ref{fppn}) are identical to those obtained in
the Zeldovich approximation; see (\ref{delc}) and (\ref{zbsol}). Therefore we
have shown that, prior to shell-crossing (i.e. while there is a one-to-one
correspondence between $\mathbf{x}$ and $\bar{\mathbf{q}}$), the \fp
approximation reduces to the \za in the semi-classical limit $\nu\rightarrow
0$.

\section{Perturbative solution of the one-dimensional \sch equation with a time-dependent potential}

We have used the wave-mechanical approach to
study the gravitational evolution of a plane-symmetric sinusoidal density perturbation in a static universe. The appropriate
one-dimensional \sch equation to be solved was

\begin{equation}
\label{a1}
\rmi\nu\frac{\partial\psi}{\partial t}=\left(-\frac{\nu^2}{2}\frac{\partial^2}{\partial x^2}+\Phi\right)\psi,
\end{equation}

\n with an external gravitational potential $\Phi=\Phi(x,t)$ of the form

\begin{equation}
\label{a3}
\Phi=D\left[\Phi_\rmi-\frac{1}{2}(D-1)v_\rmi^2\right],
\end{equation}

\n where $\Phi_\rmi=\Phi_\rmi(q)$ is the initial gravitational potential,
$v_\rmi=v_\rmi(q)$ is the initial velocity field and Eulerian
coordinates $x$ are related to Lagrangian coordinates $q$ by the Zeldovich
mapping $x=q+\tau (D-1)v_\rmi$ with $D=\exp{[(t-t_\rmi)/\tau]}$. For an initial density perturbation $\delta_\rmi=\delta_\rmi(q)$ of the form $\delta_\rmi=\delta_{\rm a}\cos{(2\pi q/d)}$ we have

\begin{equation}
\label{apigp}
\Phi_\rmi=-\left(\frac{d}{2\pi\tau}\right)^2\delta_\rmi
\end{equation}

\n and

\begin{equation}
\label{apv}
v_\rmi^2=\left(\frac{d}{2\pi\tau}\right)^2\left(\delta_{\rm a}^2-\delta_\rmi^2\right).
\end{equation}

\n We now describe how to construct a second-order perturbative
solution of the \sch equation (\ref{a1}) using TDPT. To begin with, note that (\ref{a1}) is written in the position representation. It can be written in a more general form without referring to a particular basis as 

\begin{equation}
\label{a4}
 \rmi\nu\frac{\rmd}{\rmd t}\ket{\psi(t)} = \hat{H}(t)\ket{\psi(t)},
\end{equation}

\n where $\ket{\psi(t)}$ is a general ket representing the state of the
physical system at time $t$ and $\hat{H}(t)$ is the time-dependent
total Hamiltonian. In order to solve (\ref{a4}) using PT it is convenient to split the total Hamiltonian according to $\hat{H}(t)=\hat{H}_0+\hat{\Phi}(t)$ where $\hat{H}_0=\hat{P}^2 /2$ is the free-particle Hamiltonian (with $\hat{P}$ the momentum operator) and $\hat{\Phi}(t)$ is the time-dependent potential (\ref{a3}) in operator form. Before continuing to develop a perturbative solution of (\ref{a4}) we first describe how to solve the free-particle \sch equation

\begin{equation}
\label{a7} 
 \rmi\nu\frac{\rmd}{\rmd t}\ket{\psi(t)} = \hat{H}_0\ket{\psi(t)}.
\end{equation}

\subsection{The free-particle \sch equation}
\label{fpapp}

The solution $\ket{\psi(t)}$ of the general \fp \sch equation (\ref{a7}) is obtained from the initial state ket $\ket{\psi_\rmi}=\ket{\psi(t_\rmi)}$ via

\begin{equation}
\label{a5} 
\ket{\psi(t)}=\hat{U}_0(t,t_\rmi)\ket{\psi_\rmi},
\end{equation}

\n where $\hat{U}_0(t,t_\rmi)$ is a unitary operator, known as the \fp time-evolution operator, obeying 

\begin{equation}
\label{a6} 
\rmi\nu\frac{\rmd}{\rmd t}\hat{U}_0(t,t_\rmi) = \hat{H}_0\hat{U}_0(t,t_\rmi).
\end{equation}

\n The free-particle Hamiltonian $\hat{H}_0$ is independent of time and so we can directly integrate (\ref{a6}) to find that

\begin{equation}
\label{a8} 
\hat{U}_0(t,t_\rmi) = \exp{\left[\frac{-\rmi(t-t_\rmi)\hat{H}_0}{\nu}\right]}.
\end{equation}

\n The initial state ket $\ket{\psi_\rmi}$ can be expanded in terms of
orthonormal eigenkets $\ket{n^{(0)}}$ of $\hat{H}_0$ as

\begin{equation}
\label{a9} 
\ket{\psi_\rmi} = \sum_{n} a_n\ket{n^{(0)}},
\end{equation}

\n where $a_n=\braket{n^{(0)}|\psi_\rmi}$ and the eigenkets
$\ket{n^{(0)}}$ satisfy the general time-independent free-particle \sch
equation

\begin{equation}
\label{a10} 
\hat{H}_0\ket{n^{(0)}} = E_n^{(0)}\ket{n^{(0)}}.
\end{equation}

\n In order to determine the eigenvalues $ E_n^{(0)}$ and the eigenkets
$\ket{n^{(0)}}$ we must solve the eigenvalue problem (\ref{a10}). To do this
it is convenient to first rewrite it in the position representation:

\begin{equation}
\label{a12}
-\frac{\nu^2}{2}\frac{\rmd^2 \phi_n^{(0)}}{\rmd x^2} = E_n^{(0)}\phi_n^{(0)},
\end{equation}

\n where $\phi_n^{(0)}=\phi_n^{(0)}(x)$ are the eigenfunctions, defined
by $\phi_n^{(0)}=\braket{x|n^{(0)}}$. Introducing $k_n^2=2E_n^{(0)}/\nu^2$ we find that the solutions to (\ref{a12}) are of the form

\begin{equation}
\label{a14}
\phi_n^{(0)}=\frac{1}{L^{3/2}}\exp{(\rmi k_n x)},
\end{equation}  

\n where $k_n=2n\pi/L$, $n$ an integer, since we are considering a cubic
volume of side length $L$ equipped with periodic boundary conditions at each
face. The pre-factor $1/L^{3/2}$ comes from the requirement that the
eigenfunctions are normalized.

The solution of the \fp \sch equation (\ref{a7}) follows upon inserting the
time-evolution operator (\ref{a8}) and the initial ket expansion (\ref{a9})
into (\ref{a5}). In the position representation we find

\begin{equation}
\label{a11}
\psi = \sum_{n}a_n\exp{\left[\frac{-\rmi(t-t_\rmi)E_n^{(0)}}{\nu}\right]}\phi_{n}^{(0)},
\end{equation} 

\n where $\psi=\braket{x|\psi (t)}$, $E_n^{(0)}=\nu^2 k_n^2/2$ and the
eigenfunctions $\phi_n^{(0)}$ are given by (\ref{a14}). The expansion coefficients $a_n=\braket{n^{(0)}|\psi_\rmi}$ are found from

\begin{eqnarray}
\label{an2}
a_n = L^2\int_{0}^{L}\overline{\phi_n^{(0)}}(q)\psi_{\rmi}(q)dq,
\end{eqnarray}

\n where $\psi_{\rmi}=\braket{q|\psi_\rmi}$ and an over-line denotes complex
conjugation. As discussed previously, we divide the cubic volume into cells of
side length $d$ by setting $L=Nd$, $N>0$ an integer. We can then write (\ref{an2}) as  

\begin{equation}
\label{an3}
a_n = L^2\sum_{j=0}^{N-1}\int_{jd}^{(j+1)d}\overline{\phi_n^{(0)}}(q)\psi_\rmi(q)dq.
\end{equation}

\n Performing a change of variable $q'=q-jd$ leads to

\begin{equation}
\label{an6}
a_n=L^2\int_{0}^{d}\overline{\phi_n^{(0)}}(q')\psi_{\rmi}(q')dq'\sum_{j=0}^{N-1}\exp{(-\rmi jk_n d)},
\end{equation}

\n  where we have used (\ref{a14}) along with the fact that the initial
Madelung wavefunction
$\psi_\rmi=(1+\delta_\rmi)^{1/2}\exp{(-\rmi\varphi_\rmi/\nu)}$ is periodic with period $d$. However, 

\begin{equation}
\label{an7}
\sum_{j=0}^{N-1}\exp{(-\rmi jk_n d)} = N\delta_{n,pN},
\end{equation}

\n where $p$ is an integer, and thus the coefficients $a_n$ are given by  

\begin{equation}
\label{an8}
a_n=NL^2\int_{0}^{d}\overline{\phi_n^{(0)}}(q')\psi_{\rmi}(q')dq'
\end{equation}

\n where the eigenfunctions $\phi_n^{(0)}$ are given by (\ref{a14}), but with
$k_n=2n\pi/d$. It follows from (\ref{an8}) that the expansion coefficients are simply found by taking the Fourier transform of the initial wavefunction.

\subsection{The \sch equation with a time-dependent potential}
\label{tdapp}

Our aim now is to perturbatively solve the \sch equation (\ref{a4}) with a time-dependent Hamiltonian of the form $\hat{H}(t)=\hat{H}_0+\hat{\Phi}(t)$
where the potential is given by (\ref{a3}) in the position
representation. As before, the solution $\ket{\psi(t)}$ of the \sch equation is related to the initial ket $\ket{\psi_i}$ via 

\begin{equation}
\label{b1}
\ket{\psi(t)}=\hat{U}(t,t_\rmi)\ket{\psi_\rmi},
\end{equation}

\n where the time-evolution operator obeys 

\begin{equation}
\label{b2} 
\rmi\nu\frac{\rmd}{\rmd t}\hat{U}(t,t_\rmi) = \hat{H}(t)\hat{U}(t,t_\rmi).
\end{equation}

\n The previously determined eigenkets $\ket{n^{(0)}}$ of the \fp Hamiltonian
$\hat{H}_0$ form an orthonormal set and so we may expand the initial state ket
$\ket{\psi_\rmi}$ as in (\ref{a9}) with the expansion coefficients
$a_n=\braket{n^{(0)}|\psi_{\rmi}}$ given by (\ref{an8}). In order to determine
the state ket $\ket{\psi(t)}$ at a time $t$, it is clear from (\ref{b1}) that
we require an expression for the time-evolution operator
$\hat{U}(t,t_\rmi)$. Since the Hamiltonian now depends explicitly on time, we
can no longer simply integrate (\ref{b2}) to obtain an expression for the
time-evolution operator as we did in the free-particle case. The strategy in
such a situation is to seek an approximate expression for the time-evolution operator. We now describe how this may be achieved by using TDPT.

\subsubsection{Time-dependent perturbation theory}

In this work we have exclusively used a description of quantum
dynamics known as the \sch picture. However, TDPT solutions of the \sch
equation with a time-dependent potential are most easily constructed by
appealing to another equivalent description of quantum dynamics known as the interaction picture. A general state ket in the interaction picture $\ket{\psi(t)}^{(\rm I)}$ is related to the state ket in the \sch picture $\ket{\psi(t)}$ by

\begin{equation}
\label{b7}
\ket{\psi(t)}^{(\rm I)} =\hat{U}_{0}^{\dagger}(t,t_\rmi)\ket{\psi(t)},
\end{equation}

\n where the free-particle time-evolution operator $\hat{U}_{0}(t,t_\rmi)$ is
given by (\ref{a8}) and a dagger denotes the adjoint. The superscript $(\rm
I)$ will be used throughout to denote quantities in the interaction
picture. Note that $\ket{\psi_\rmi}^{(\rm I)}=\ket{\psi_\rmi}$ and so the
interaction and \sch picture state kets initially coincide. In the interaction picture it is straightforward to show that a general state ket evolves according to:

\begin{equation}
\label{b9}
\rmi\nu\frac{\rmd}{\rmd t}\ket{\psi(t)}^{(\rm I)}=\hat{\Phi}^{(\rm
  I)}(t)\ket{\psi(t)}^{(\rm I)},
\end{equation}

\n where $\hat{\Phi}^{(\rm
  I)}(t)\equiv\hat{U}_{0}^{\dagger}(t,t_\rmi)\hat{\Phi}(t)\hat{U}_{0}(t,t_\rmi)$ is the perturbing potential in the interaction picture. It is apparent from (\ref{b9}) that the time-evolution of a state ket in the interaction picture is determined solely by $\hat{\Phi}^{(\rm I)}(t)$. 

The objective is to determine a perturbation expansion for the time-evolution
operator $\hat{U}(t,t_\rmi)$ in the \sch picture. This is best achieved by
first finding a perturbation expansion for the time-evolution operator in the
interaction picture. In this picture a time-evolution operator $\hat{U}^{(\rm I)}(t,t_\rmi)$ can be defined by 

\begin{equation}
\label{b10}
\ket{\psi(t)}^{(\rm I)} = \hat{U}^{(\rm I)}(t,t_\rmi)\ket{\psi_\rmi}^{(\rm I)}.
\end{equation}

\n Substituting (\ref{b10}) into (\ref{b9}) it is clear that $\hat{U}^{(\rm I)}(t,t_\rmi)$ obeys

\begin{equation}
\label{b11}
\rmi\nu\frac{\rmd}{\rmd t}\hat{U}^{(\rm I)}(t,t_\rmi) = \hat{\Phi}^{(\rm
  I)}(t)\hat{U}^{(\rm I)}(t,t_\rmi),
\end{equation}

\n where this equation must be solved subject to the initial
condition $\hat{U}^{(\rm I)}(t_\rmi,t_\rmi) = I$ with $I$ the identity
operator. Observe that equation (\ref{b11}), together with the appropriate initial condition, is equivalent to the integral equation

\begin{equation}
\label{b12}
\hat{U}^{(\rm I)}(t,t_\rmi) =
I-\frac{\rmi}{\nu}\int_{t_\rmi}^{t}\hat{\Phi}^{(\rm I)}(t')\hat{U}^{(\rm I)}(t',t_\rmi)dt'.
\end{equation}

\n The integral equation (\ref{b12}) provides a convenient means of
determining a perturbation expansion for $\hat{U}^{(\rm I)}(t,t_\rmi)$. By
iteration we find that

\begin{equation}
\label{b13}
\hat{U}^{(\rm I)}(t,t_\rmi) = I -
\frac{\rmi}{\nu}\int_{t_\rmi}^{t}dt'\hat{\Phi}^{(\rm
  I)}(t')-\frac{1}{\nu^2}\int_{t_\rmi}^{t}dt'\int_{t_\rmi}^{t'}dt''\hat{\Phi}^{(\rm I)}(t')\hat{\Phi}^{(\rm I)}(t''),
\end{equation}

\n to second-order. To find the corresponding time-evolution operator
$\hat{U}(t,t_\rmi)$ in the \sch picture, first note that $\ket{\psi(t)}^{(\rm I)} = \hat{U}^{(\rm I)}(t,t_\rmi)\ket{\psi_\rmi}^{(\rm I)}=\hat{U}^{(\rm I)}(t,t_\rmi)\ket{\psi_\rmi}$ since the state kets in the interaction and \sch pictures coincide at $t=t_\rmi$. Using the definition (\ref{b7}) it follows that

\begin{equation}
\label{b14}
\ket{\psi(t)} = \hat{U}_{0}(t,t_\rmi)\hat{U}^{(\rm I)}(t,t_\rmi)\ket{\psi_\rmi}.
\end{equation}

\n Upon comparison with (\ref{b1}) it is immediately clear that the
time-evolution operator in the \sch picture $\hat{U}(t,t_\rmi)$ is related to
$\hat{U}^{(\rm I)}(t,t_\rmi)$ via

\begin{equation}
\label{b15}
\hat{U}(t,t_\rmi) = \hat{U}_{0}(t,t_\rmi)\hat{U}^{(\rm I)}(t,t_\rmi).
\end{equation}

\n Multiplying (\ref{b13}) by $\hat{U}_{0}(t,t_\rmi)$, inserting
$\hat{\Phi}^{(\rm
  I)}(t)=\hat{U}_{0}^{\dagger}(t,t_\rmi)\hat{\Phi}(t)\hat{U}_{0}(t,t_\rmi)$
and using the following property of the time-evolution operator:
$\hat{U}_{0}(t,t_\rmi)\hat{U}_{0}^{\dagger}(t',t_\rmi)=\hat{U}_{0}(t,t_\rmi)\hat{U}_{0}(t_\rmi,t')=\hat{U}_{0}(t,t')$
then gives

\begin{equation}
\label{b16}
\hat{U}(t,t_\rmi) = \sum_{j=0}^2\hat{U}_{j}(t,t_\rmi),
\end{equation}

\n to second order, where $\hat{U}_{0}(t,t_\rmi)$ is given by (\ref{a8}),

\begin{equation}
\label{b17}
\hat{U}_{1}(t,t_\rmi)  = -\frac{\rmi}{\nu}\int_{t_\rmi}^{t}dt'\hat{U}_{0}(t,t')\hat{\Phi}(t')\hat{U}_{0}(t',t_\rmi),
\end{equation}

\n and

\begin{equation}
\label{b18}
\hat{U}_{2}(t,t_\rmi) = -\frac{1}{\nu^2}\int_{t_\rmi}^{t}dt'\int_{t_\rmi}^{t'}dt''\hat{U}_{0}(t,t')\hat{\Phi}(t')\hat{U}_{0}(t',t'')\hat{\Phi}(t'')\hat{U}_{0}(t'',t_\rmi).
\end{equation}

\n The second-order expression for the time-evolution operator (\ref{b16}) can
be inserted into (\ref{b1}) along with the initial ket expansion (\ref{a9}) to
give the following approximate solution of the \sch equation (\ref{a4}):

\begin{equation}
\label{b19}
\ket{\psi(t)}=\sum_{j=0}^2\ket{\psi^{(j)}(t)},
\end{equation}

\n where 

\begin{equation}
\label{bn1}
\ket{\psi^{(j)}(t)}=\sum_{n}a_n\hat{U}_{j}(t,t_\rmi)\ket{n^{(0)}},
\end{equation}

\n with $\hat{U}_0(t,t_\rmi)$, $\hat{U}_{1}(t,t_\rmi)$ and
$\hat{U}_{2}(t,t_\rmi)$ given by (\ref{a8}), (\ref{b17}) and (\ref{b18}),
respectively. In the position representation the zeroth-order term
$\psi^{(0)}=\braket{x|\psi^{(0)}(t)}$ is simply the solution (\ref{a11}) of
the \fp \sch equation discussed earlier. Since the eigenfunctions
$\phi_n^{(0)}$ of the \fp Hamiltonian are independent of time, we may write the first-order term $\psi^{(1)}=\braket{x|\psi^{(1)}(t)}$ as

\begin{eqnarray}
\label{b21}
\psi^{(1)} = -\frac{\rmi}{\nu}\sum_{n}a_n\sum_{m}\mathcal{I}_{m,n}\phi_m^{(0)},
\end{eqnarray}

\n where $\mathcal{I}_{m,n}=\mathcal{I}_{m,n}(t)$ is defined by 

\begin{equation}
\mathcal{I}_{m,n}=\int_{t_\rmi}^{t}dt'\exp{\left[\frac{-\rmi (t-t')E_m^{(0)}}{\nu}\right]}\Phi_{m,n}(t')\exp{\left[\frac{-\rmi (t'-t_\rmi)E_{n}^{(0)}}{\nu}\right]},
\end{equation}

\n and the matrix elements $\Phi_{m,n}(t)=\braket{m^{(0)}|\hat{\Phi}(t)|n^{(0)}}$ are given by

\begin{equation}
\label{b22}
\Phi_{m,n}= L^2\int_{0}^{L}\overline{\phi_m^{(0)}}(x)\Phi(x,t')\phi_n^{(0)}(x)dx.
\end{equation}

\n In order to calculate the matrix elements $\Phi_{m,n}$, we first use the \za
$x=q+\tau (D-1)v_\rmi$ to change to Lagrangian coordinates $q$. Inserting
$\Phi$ from (\ref{a3}) and using (\ref{a14}) then leads to

\begin{eqnarray}
\fl\Phi_{m,n}=\frac{1}{L}\int_{0}^{L}D\left[\Phi_\rmi(q)-\frac{1}{2}(D-1)v_\rmi^2(q)\right]\exp{(-ik_p q)}\exp{\left[-ik_p\tau (D-1)v_\rmi(q)\right]}\nonumber\\
\times\mathcal{J}(q,t')dq\label{b23},
\end{eqnarray}

\n where $k_p=k_m-k_n$ and the Jacobian determinant $\mathcal{J}=\partial
x/\partial q=1+\tau (D-1)\rmd v_\rmi/\rmd q$. Upon substituting $\Phi_\rmi$ and $v_\rmi$
from (\ref{apigp}) and (\ref{apv}) we obtain, after a lengthy calculation,  

\begin{eqnarray}
\label{matel}
\fl\Phi_{m,n}=\frac{\delta_{\rm a}}{2}\left(\frac{d}{2\pi\tau}\right)^2
D\left\{\alpha\left[J_p(2p\alpha)+\frac{3}{2}\sum_{s=\pm 2}J_{p+s}(2p\alpha)\right]+\left(\frac{\alpha^2}{2}-1\right)\sum_{s=\pm 1}J_{p+s}(2p\alpha)\right.\nonumber\\
\left.-\frac{\alpha^2}{2}\sum_{s=\pm 3}J_{p+s}(2p\alpha)\right\},
\end{eqnarray}

\n where $p=m-n$ is an integer, $\alpha=\alpha(t)$ is defined by $\alpha=\delta_{\rm a}(D-1)/2$ and

\begin{equation}
\label{bess}
J_l(x)=\frac{1}{2\pi}\int_{-\pi}^{\pi}\exp{(il\theta)}\exp{[-ix\sin{(\theta)}]}d\theta
\end{equation}

\n are Bessel functions of the first kind. In a similar manner we find that
the second-order term $\psi^{(2)}=\braket{x|\psi^{(2)}(t)}$ is given by

\begin{eqnarray}
\psi^{(2)} = -\frac{1}{\nu^2}\sum_{n}a_n\sum_{m}\sum_{l}\mathcal{K}_{l,m,n}\phi_l^{(0)},
\end{eqnarray}

\n where $\mathcal{K}_{l,m,n}=\mathcal{K}_{l,m,n}(t)$ is defined by

\begin{eqnarray}
\fl\mathcal{K}_{l,m,n}=\int_{t_\rmi}^{t}dt'\int_{t_\rmi}^{t'}dt''\exp{\left[\frac{-\rmi (t-t')E_l^{(0)}}{\nu}\right]}\Phi_{l,m}(t')\exp{\left[\frac{-\rmi (t'-t'')E_{m}^{(0)}}{\nu}\right]}\nonumber\\
\times\Phi_{m,n}(t'')\exp{\left[\frac{-\rmi (t''-t_\rmi)E_{n}^{(0)}}{\nu}\right]},
\end{eqnarray}

\n and the matrix elements
$\Phi_{l,m}(t)=\braket{l^{(0)}|\hat{\Phi}(t)|m^{(0)}}$ and
$\Phi_{m,n}(t)=\braket{m^{(0)}|\hat{\Phi}(t)|n^{(0)}}$ are obtained from (\ref{matel}).

\end{document}